\pdfoutput=1
\documentclass[12pt,a4paper]{article}

\usepackage{ifthen} 
\newboolean{pdflatex}
\setboolean{pdflatex}{true} 

\newboolean{articletitles}
\setboolean{articletitles}{true} 

\newboolean{uprightparticles}
\setboolean{uprightparticles}{false} 

\def\paperauthors{LHCb collaboration} 
\def\paperasciititle{Deuteron identification via time of flight with LHCb} 
\def\papertitle{Deuteron identification via time of flight with \mbox{LHCb}\xspace} 
\def\paperkeywords{{High Energy Physics}, {LHCb}} 
\def\papercopyright{\the\year\ CERN for the benefit of the LHCb collaboration} 
\def\paperlicence{CC BY 4.0 licence}
\def\paperlicenceurl{https://creativecommons.org/licenses/by/4.0/}

\newif\ifEnableSectionTOCLinks
\EnableSectionTOCLinksfalse 

\usepackage[top=1in, bottom=1.25in, left=1in, right=1in]{geometry}
\usepackage{microtype}
\usepackage{lineno}  
\usepackage{xspace} 
\usepackage{caption} 

\usepackage{graphicx}  
\usepackage{color}
\usepackage{colortbl}
\graphicspath{{./figs/}} 

\usepackage{amsmath} 
\usepackage{amssymb}
\usepackage{amsfonts}
\usepackage{upgreek} 

\newcommand*\patchAmsMathEnvironmentForLineno[1]{%
\expandafter\let\csname old#1\expandafter\endcsname\csname #1\endcsname
\expandafter\let\csname oldend#1\expandafter\endcsname\csname
end#1\endcsname
 \renewenvironment{#1}%
   {\linenomath\csname old#1\endcsname}%
   {\csname oldend#1\endcsname\endlinenomath}%
}
\newcommand*\patchBothAmsMathEnvironmentsForLineno[1]{%
  \patchAmsMathEnvironmentForLineno{#1}%
  \patchAmsMathEnvironmentForLineno{#1*}%
}
\AtBeginDocument{%
\patchBothAmsMathEnvironmentsForLineno{equation}%
\patchBothAmsMathEnvironmentsForLineno{align}%
\patchBothAmsMathEnvironmentsForLineno{flalign}%
\patchBothAmsMathEnvironmentsForLineno{alignat}%
\patchBothAmsMathEnvironmentsForLineno{gather}%
\patchBothAmsMathEnvironmentsForLineno{multline}%
\patchBothAmsMathEnvironmentsForLineno{eqnarray}%
}

\usepackage[pdftex,
            pdfauthor={\paperauthors},
            pdftitle={\paperasciititle},
            pdfkeywords={\paperkeywords}]{hyperref}
\usepackage{hyperxmp}
\hypersetup{
    pdfcopyright={Copyright (C) \papercopyright},
    pdflicenseurl={\paperlicenceurl}
}

\usepackage[colorinlistoftodos,textsize=scriptsize]{todonotes}

\usepackage[bottom,flushmargin,hang,multiple]{footmisc}

\usepackage[all]{hypcap} 

\usepackage{xspace}
\usepackage{upgreek}

\def\lhcb   {\mbox{LHCb}\xspace}

\def\MagUp {\mbox{\em Mag\kern -0.05em Up}\xspace}

\ifthenelse{\boolean{uprightparticles}}%
{

 \def\Ppi         {\ensuremath{\uppi}\xspace}

 \def\PDelta      {\ensuremath{\Delta}\xspace}
 \def\PXi         {\ensuremath{\Xi}\xspace}
 \def\PLambda     {\ensuremath{\Lambda}\xspace}
 \def\PSigma      {\ensuremath{\Sigma}\xspace}
 \def\POmega      {\ensuremath{\Omega}\xspace}
 \def\PUpsilon    {\ensuremath{\Upsilon}\xspace}
 \let\oldPi\Pi
 \def\PPi         {\ensuremath{\oldPi}\xspace}

 \def\PB      {\ensuremath{\mathrm{B}}\xspace}
 \def\PD      {\ensuremath{\mathrm{D}}\xspace}

 \def\PK      {\ensuremath{\mathrm{K}}\xspace}
 \def\Pp      {\ensuremath{\mathrm{p}}\xspace}

 \def\Ps      {\ensuremath{\mathrm{s}}\xspace}

 \def\thebaroffset{0.0em}
}
{

 \def\Ppi         {\ensuremath{\pi}\xspace}

 \mathchardef\PDelta="7101
 \mathchardef\PXi="7104
 \mathchardef\PLambda="7103
 \mathchardef\PSigma="7106
 \mathchardef\POmega="710A
 \mathchardef\PUpsilon="7107
 \mathchardef\PPi="7105
 \def\PB      {\ensuremath{B}\xspace}
 \def\PD      {\ensuremath{D}\xspace}

 \def\PK      {\ensuremath{K}\xspace}
 \def\Pp      {\ensuremath{p}\xspace}

 \def\Ps      {\ensuremath{s}\xspace}

 \def\thebaroffset{0.18em}
}
\newcommand{\offsetoverline}[2][\thebaroffset]{\kern #1\overline{\kern -#1 #2}}%

\makeatletter
\ifcase \@ptsize \relax
  \newcommand{\miniscule}{\@setfontsize\miniscule{4}{5}}
\or
  \newcommand{\miniscule}{\@setfontsize\miniscule{5}{6}}
\or
  \newcommand{\miniscule}{\@setfontsize\miniscule{5}{6}}
\fi
\makeatother

\DeclareRobustCommand{\optbar}[1]{\shortstack{{\miniscule (\rule[.5ex]{1.25em}{.18mm})}
  \\ [-.7ex] $#1$}}

\def\squark    {{\ensuremath{\Ps}}\xspace}

\def\pion   {{\ensuremath{\Ppi}}\xspace}

\def\pip    {{\ensuremath{\pion^+}}\xspace}
\def\pim    {{\ensuremath{\pion^-}}\xspace}

\def\kaon    {{\ensuremath{\PK}}\xspace}

\def\KorKbar {\kern \thebaroffset\optbar{\kern -\thebaroffset \PK}{}\xspace}

\def\KS      {{\ensuremath{\kaon^0_{\mathrm{S}}}}\xspace}

\def\D       {{\ensuremath{\PD}}\xspace}

\def\DorDbar {\kern \thebaroffset\optbar{\kern -\thebaroffset \PD}\xspace}

\def\Dp      {{\ensuremath{\D^+}}\xspace}
\def\Dm      {{\ensuremath{\D^-}}\xspace}

\def\DpDm    {\ensuremath{\Dp {\kern -0.16em \Dm}}\xspace}

\def\B       {{\ensuremath{\PB}}\xspace}

\def\BorBbar {\kern \thebaroffset\optbar{\kern -\thebaroffset \PB}\xspace}

\def\Bd      {{\ensuremath{\B^0}}\xspace}

\def\BdorBdbar {\kern \thebaroffset\optbar{\kern -\thebaroffset \Bd}\xspace}

\def\Bs      {{\ensuremath{\B^0_\squark}}\xspace}

\def\BsorBsbar {\kern \thebaroffset\optbar{\kern -\thebaroffset \Bs}\xspace}

\def\Y#1S{\ensuremath{\PUpsilon{(#1S)}}\xspace}

\def\proton      {{\ensuremath{\Pp}}\xspace}

\def\Lz          {{\ensuremath{\PLambda}}\xspace}

\def\LorLbar     {\kern \thebaroffset\optbar{\kern -\thebaroffset \PLambda}\xspace}

\newcommand{\decay}[2]{\mbox{\ensuremath{#1\!\to #2}}\xspace}

\def\to                 {\ensuremath{\rightarrow}\xspace}

\def\order   {{\ensuremath{\mathcal{O}}}\xspace}

\def\AT#1     {\ensuremath{A_{\mathrm{T}}^{#1}}\xspace}           

\def\C#1      {\ensuremath{\mathcal{C}_{#1}}\xspace}                       
\def\Cp#1     {\ensuremath{\mathcal{C}_{#1}^{'}}\xspace}                    
\def\Ceff#1   {\ensuremath{\mathcal{C}_{#1}^{\mathrm{(eff)}}}\xspace}        
\def\Cpeff#1  {\ensuremath{\mathcal{C}_{#1}^{'\mathrm{(eff)}}}\xspace}       
\def\Ope#1    {\ensuremath{\mathcal{O}_{#1}}\xspace}                       
\def\Opep#1   {\ensuremath{\mathcal{O}_{#1}^{'}}\xspace}                    

\newcommand{\aunit}[1]{\ensuremath{\text{\,#1}}}

\newcommand{\tev}{\aunit{Te\kern -0.1em V}\xspace}
\newcommand{\gev}{\aunit{Ge\kern -0.1em V}\xspace}
\newcommand{\mev}{\aunit{Me\kern -0.1em V}\xspace}
\newcommand{\kev}{\aunit{ke\kern -0.1em V}\xspace}
\newcommand{\ev}{\aunit{e\kern -0.1em V}\xspace}

\newcommand{\mevc}{\ensuremath{\aunit{Me\kern -0.1em V\!/}c}\xspace}
\newcommand{\gevc}{\ensuremath{\aunit{Ge\kern -0.1em V\!/}c}\xspace}
\newcommand{\mevcc}{\ensuremath{\aunit{Me\kern -0.1em V\!/}c^2}\xspace}
\newcommand{\gevcc}{\ensuremath{\aunit{Ge\kern -0.1em V\!/}c^2}\xspace}

\def\mm   {\aunit{mm}\xspace}

\def\nb {\aunit{nb}\xspace}
\def\invnb {\ensuremath{\nb^{-1}}\xspace}

\def\ns   {\ensuremath{\aunit{ns}}\xspace}

\def\order{{\ensuremath{\mathcal{O}}}\xspace}
\newcommand{\chisq}{\ensuremath{\chi^2}\xspace}

\def\gsim{{~\raise.15em\hbox{$>$}\kern-.85em
          \lower.35em\hbox{$\sim$}~}\xspace}
\def\lsim{{~\raise.15em\hbox{$<$}\kern-.85em
          \lower.35em\hbox{$\sim$}~}\xspace}

\def\sqsnn {\ensuremath{\protect\sqrt{s_{\scriptscriptstyle\text{NN}}}}\xspace}

\def\geant      {\mbox{\textsc{Geant4}}\xspace}

\def\pythia     {\mbox{\textsc{Pythia}}\xspace}

\def\tensorflow {\mbox{\textsc{TensorFlow}}\xspace}

\def\tell1  {TELL1\xspace}
\def\ukl1   {UKL1\xspace}

\newcommand{\lhcborcid}[1]{\href{https://orcid.org/#1}{\hspace*{0.1em}\raisebox{-0.45ex}{\includegraphics[width=1em]{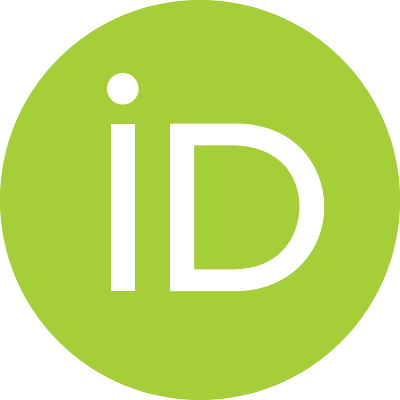}}}}

\hypersetup{
  colorlinks   = true, 
  urlcolor     = blue, 
  linkcolor    = blue, 
  citecolor    = red   
}

\ifEnableSectionTOCLinks
    \usepackage[explicit]{titlesec} 
    
    \let\oldcontentsline\contentsline
    \renewcommand

    \titleformat{\section}{\normalfont\Large\bf}{\hyperlink{tocsection.\thesection}{{\thesection} \parbox[t]{\dimexpr\textwidth-1pc}{#1}}}{1pc}{}

    \titleformat{\subsection}{\normalfont\bf}{\hyperlink{tocsubsection.\thesubsection}{{\thesubsection} \parbox[t]{\dimexpr\textwidth-1pc}{#1}}}{1pc}{}

    \titleformat{name=\section,numberless}[display]{}{}{0pt}{\normalfont\Huge\bfseries #1}
\fi

\usepackage{cite} 
\usepackage{mciteplus}

\usepackage{subcaption}

\def\sqsnn{\ensuremath{\protect\sqrt{s_{\scriptscriptstyle\rm NN}}}\xspace}
\def\pHe{\ensuremath{\proton\text{He}}\xspace}

\def\pPb{\ensuremath{\proton\text{Pb}}\xspace}

\def\hatbeta{\ensuremath{\hat{\beta}}\xspace}
\def\hatbetaNN{\ensuremath{\hat{\beta}_{\text{ANN}}}\xspace}
\def\hatM{\ensuremath{\hat{\rm M}_{\text{TOF}}}\xspace}

\def\dbeta{\ensuremath{\Delta\beta}\xspace}
\def\dbetaNN{\ensuremath{\Delta\beta_{\text{ANN}}}\xspace}

\usepackage{longtable} 

\begin{document}

\renewcommand{\thefootnote}{\fnsymbol{footnote}}
\setcounter{footnote}{1}

\begin{titlepage}
\pagenumbering{roman}

\vspace*{-1.5cm}
\centerline{\large EUROPEAN ORGANIZATION FOR NUCLEAR RESEARCH (CERN)}
\vspace*{1.5cm}
\noindent
\begin{tabular*}{\linewidth}{lc@{\extracolsep{\fill}}r@{\extracolsep{0pt}}}
\vspace*{-1.5cm}\mbox{\!\!\!\includegraphics[width=.14\textwidth]{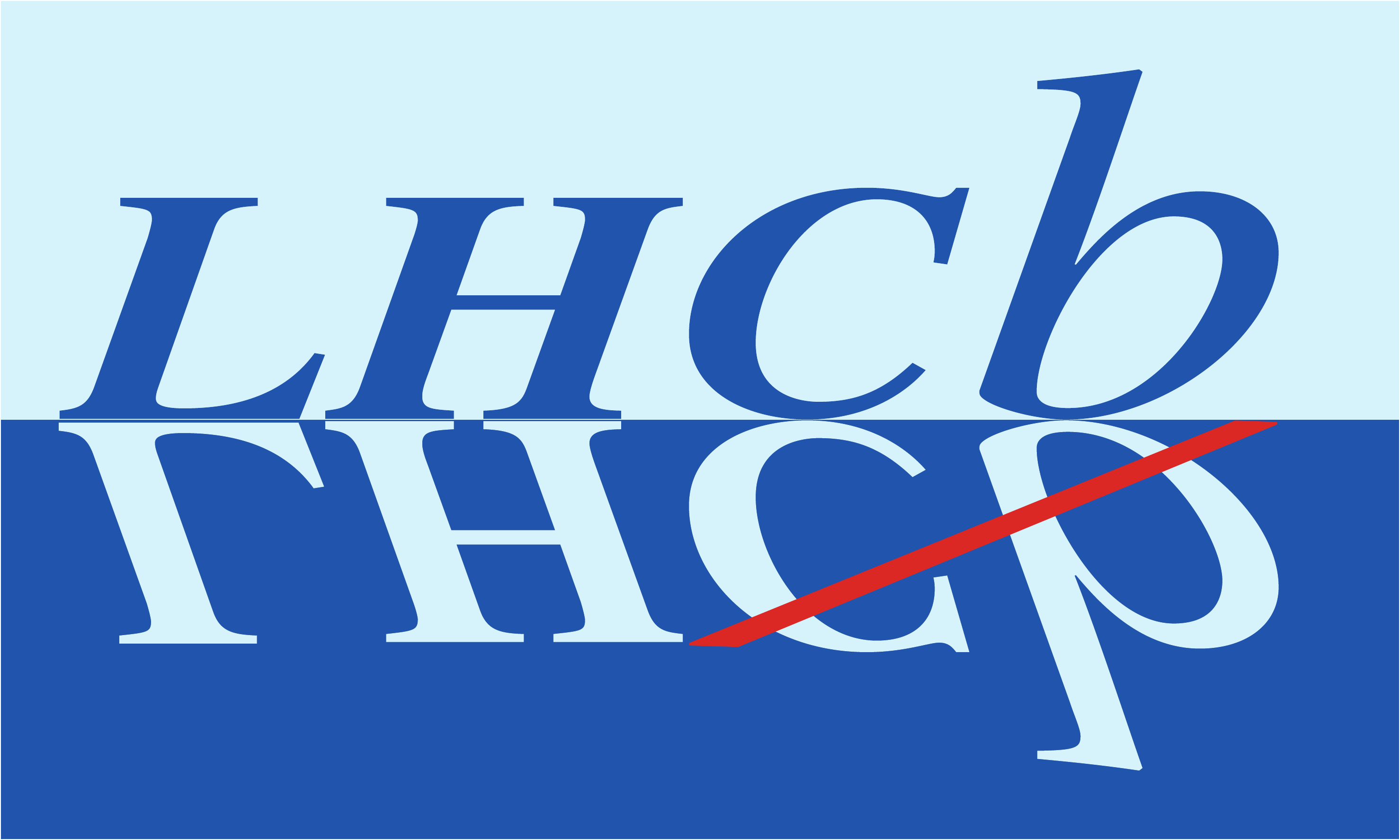}}
\\
 & & LHCb-DP-2025-004 \\  
 & & November 19, 2025 \\ 
\end{tabular*}

\vspace*{4.0cm}

{\normalfont\bfseries\boldmath\huge
\begin{center}
  \papertitle 
\end{center}
}

\vspace*{2.0cm}

\begin{center}
\paperauthors\footnote{Authors are listed at the end of this paper.}
\end{center}

\vspace{\fill}

\begin{abstract}
  \noindent
  It is shown that the timing capabilities of the \mbox{LHCb}\xspace detector operated during the LHC Run~2
  can be used to identify light ion particles with momenta of a few\,\ensuremath{\aunit{Ge\kern -0.1em V\!/}c}\xspace.
  This is achieved by estimating the particle time of flight through a newly developed technique. A dedicated reconstruction procedure and a neural-network-based estimator of the particle speed have been developed to enable deuteron identification by suppressing the abundant background from lighter particles.  
  The performance of the identification procedure is demonstrated in a sample 
  of proton-helium collisions at \ensuremath{\protect\sqrt{s_{\scriptscriptstyle\text{NN}}}}\xspace=110\aunit{Ge\kern -0.1em V}\xspace,
  where the production of deuteron and triton particles is observed. 
  This novel approach opens the way to study deuteron and antideuteron production
  for different collision systems at different energy scales, exploiting the rich dataset collected by the \mbox{LHCb}\xspace experiment.

\end{abstract}

\vspace*{2.0cm}

\begin{center}
  Published in The European Physical Journal C 85 (2025) 1329
\end{center}

\vspace{\fill}

{\footnotesize 
\centerline{\copyright~\papercopyright. \href{\paperlicenceurl}{\paperlicence}.}}
\vspace*{2mm}

\end{titlepage}


\newpage
\setcounter{page}{2}
\mbox{~}

\renewcommand{\thefootnote}{\arabic{footnote}}
\setcounter{footnote}{0}


\cleardoublepage


\pagestyle{plain} 
\setcounter{page}{1}
\pagenumbering{arabic}


\section{Introduction}
    
The measurement of antimatter in cosmic rays is a sensitive probe of
physics beyond the Standard Model, as dark matter annihilation or
decay could contribute to the flux of cosmic antiparticles~\cite{Salati:2010rc}.
The AMS-02 experiment~\cite{AMS:2021nhj} 
has reported several antideuteron and antihelium candidates whose
origin is still unclear~\cite{STingCERNseminar}. These particles could
be produced in collisions between
cosmic rays and the interstellar medium via coalescence of secondary
antibaryons.  While measurements of the related production cross
sections have been performed at RHIC and LHC hadron colliders for a
nucleon-nucleon centre-of-mass energy \sqsnn well above 100\gev, 
very limited data exist for the energy scale relevant to the AMS-02
observations, namely \sqsnn between $10$ and $100$\gev.

The \lhcb detector~\cite{LHCb-DP-2008-001,LHCb-DP-2014-002} is a
single-arm forward spectrometer covering the pseudorapidity range $2<\eta<5$, designed for
the study of heavy-flavour particles in proton-proton ($pp$) collisions.
During the LHC Run 2, it was also operated with \pPb and PbPb
collisions. Production studies in beam-beam collisions at the \lhcb experiment can
complement those from the other LHC detectors, which mostly cover
central rapidity.
Moreover, thanks to the SMOG gas target~\cite{smog}, the \lhcb experiment also
collected a rich dataset of fixed-target collisions of proton and
lead beams on He, Ne and Ar targets, at \sqsnn between 68 and 110\gev~\cite{Bursche:2649878}.

The detector includes two ring-imaging Cherenkov (RICH) detectors that
are able to discriminate between different species of charged hadrons,
while leptons, photons and neutral hadrons are identified by a
calorimeter system and a muon system consisting of five stations.
Though light ion identification was not among the requirements of the
detector design, the AMS-02 observation motivated an effort to
identify light nuclei from the available detector information.
In particular, a recent study has demonstrated that helium nuclei can be identified via their characteristic energy loss in the tracking detectors~\cite{LHCb-DP-2023-002}. This recent development opens a promising avenue for light ion identification at LHCb, despite not being part of the initial design goals.

Active deuteron\footnote{The inclusion of charge-conjugated particles is implied in the following, unless otherwise stated.} identification in the RICH detectors is possible only 
for momenta above 35\gevc and is limited by the abundant pion
background~\cite{Baker:2019fjx}. In this paper, the identification of low-momentum 
deuteron particles through a time-of-flight measurement is studied.

The \lhcb Outer Tracker (OT)~\cite{LHCb-TDR-006} consists of 12
double layers of straw tubes, arranged in three stations placed
downstream of the spectrometer's magnet, at a distance between 7.9~m and
9.4~m from the nominal beam-beam collision point. 
Its large area covers the outer region of the spectrometer's acceptance,
corresponding to particles of lower rapidity and momentum.
For the LHC Run 2, the detector was precisely recalibrated, reaching
an average resolution on the drift time of 2.4 ns per hit~\cite{LHCb-DP-2017-001}. 
For an average number of 18 OT hits per track, 
the resolution for the time-of-flight measurement is therefore 0.57 ns, 
allowing, in principle, to distinguish the proton and deuteron hypotheses 
by more than 3 standard deviations for momenta below 4.6\gevc.
The first muon station (M1), located between the OT and the calorimeters,
detects charged particles with an efficiency above 98.5\% and
provides an additional independent measurement of the particle time of flight with 
a resolution below 4 ns~\cite{LHCb-DP-2012-002}. The timing capabilities of the OT and of M1 detectors form the basis of the technique presented in this paper. 

The paper is organised as follows: a dedicated track reconstruction procedure in
the OT, needed to efficiently reconstruct the deuteron
candidates, is presented in Section~\ref{sec:reco}.
The strategy used to determine the particle velocity and suppress
the background to deuteron identification is discussed in
Section~\ref{sec:idtool}. In Section~\ref{sec:performance}, the achieved
performance of the identification procedure is presented on a sample of
\pHe collisions collected at \mbox{\sqsnn=110\gev}. Finally, in Section~\ref{sec:conclusions} the conclusions summarizing the performance and general applicability of the proposed method are presented.

\section{Deuteron reconstruction}
\label{sec:reco}

In the standard track reconstruction procedure used in the \lhcb experiment (see Ref.~\cite{DeCianThesis} for a detailed description),
candidates for particles produced at the primary collision and crossing
the \lhcb spectrometer are obtained through a pattern recognition algorithm 
known as Forward Tracking~\cite{PatForward}. The algorithm 
extrapolates track segments reconstructed in the vertex detector and
searches for detector hits in the other tracking stations.
Hits in the OT are obtained by combining the known wire position of each
OT straw tube with the measured drift time of the signals. The drift
times are determined after taking into account the 
particle time of flight according to its trajectory from the nominal
beam-beam collision point to the hit position, assuming that the particle
travels at the speed of light $c$. For particles travelling at
lower speed ($\beta=v/c < 1$), the delay of the particle arrival at the OT with respect to the
ultrarelavistic hypothesis biases the reconstructed
drift times.  It has been shown in
Ref.~\cite{LHCb-DP-2017-001} that  the average time residual 
of OT hits with respect to the
fitted track provides some 
discrimination power between protons and pions. 
The overestimation of the drift time for particles with $\beta \ll 1$ produces a bias to their reconstructed positions in the OT. Moreover, in the standard
reconstruction, late OT hits are discarded if they are compatible
with being produced by a collision following the acquired one by 25~ns, given the nominal 40 MHz collision rate.
The quality of the reconstructed tracks for such ``slow'' particles is
therefore degraded, possibly causing the track
candidates  to be rejected.

To evaluate this effect and benchmark an improved deuteron reconstruction, three simulation
samples are produced. In the first sample, single deuteron 
particles with momentum between 1 and 15\gevc are simulated in the detector. 
In the other samples, a more realistic simulation of \pHe collisions at
\sqsnn=110\gev is performed, requiring the presence of a deuteron in
the final state.  Collisions are generated using two different models,
 QGSJETII04~\cite{Ostapchenko:2010vb} or
\pythia 6.4~\cite{Sjostrand:2006za}, providing different 
predictions for the baryon production. Only the former includes baryon production 
through fragmentation of the target He nucleus. Deuteron production via coalescence 
of produced baryons is not included in either model and is simulated through 
an additional custom algorithm, similar to that described in  Ref.~\cite{ALICE-PUBLIC-2017-010}.
The interaction of the generated particles with the detector, and its
response, are implemented using the \geant
toolkit~\cite{Allison:2006ve, *Agostinelli:2002hh} as described in
Ref.~\cite{LHCb-PROC-2011-006}.

\begin{figure}[t]
	\centering
    \includegraphics[width=0.6\textwidth]{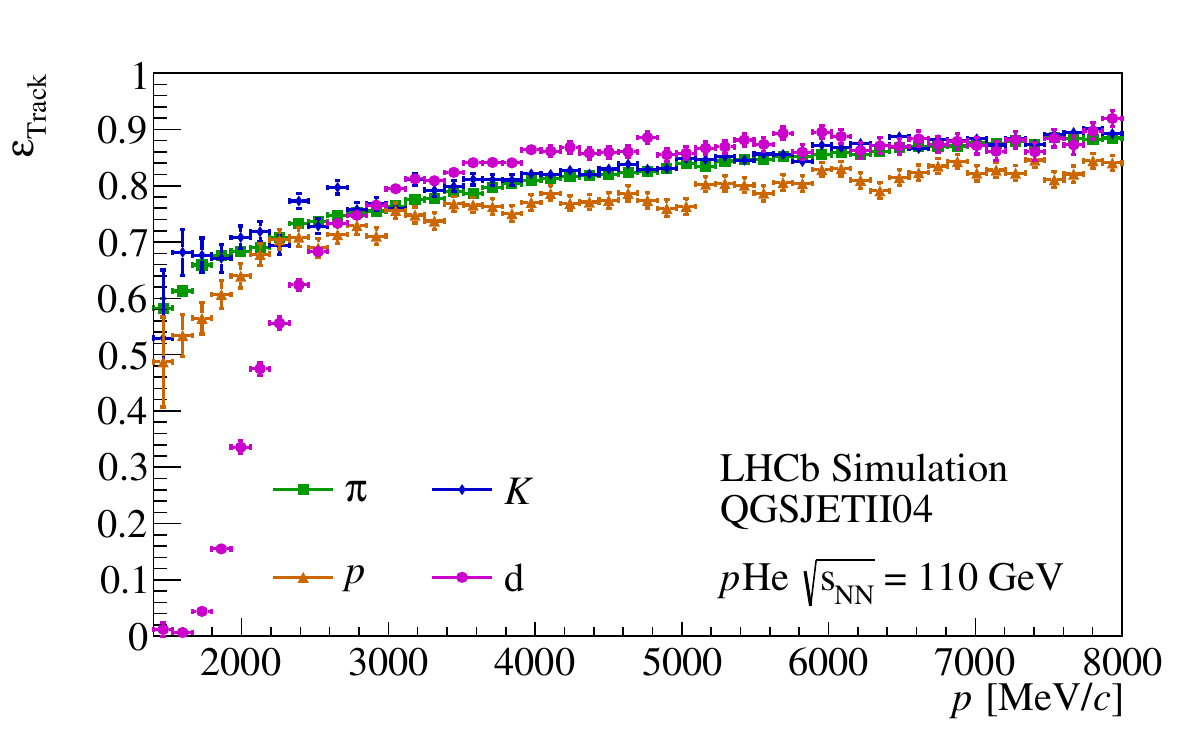}
	\caption{Track reconstruction efficiency $\varepsilon_{\rm Track}$ of the standard pattern recognition algorithm as a function of the momentum \textit{p} for different hadron species in the QGSJET-based simulation sample.} 
	\label{fig:effForward}
\end{figure}

The efficiency of the standard \lhcb pattern recognition algorithm for
different hadron species in simulated events is shown
in Fig.~\ref{fig:effForward}. It is computed for particles in the detector acceptance that produce enough detector hits to be possibly reconstructed. The efficiency for deuterons is strongly degraded for momenta
below 3\gevc, namely in the most promising kinematic region for
deuteron identification via time of flight.

A modified version of the pattern recognition algorithm, named
TofForward  hereafter, is developed to recover the slow deuteron candidates. All recorded OT hits are considered, including late signals
that could be produced by low-momentum deuterons.
In an initial step, the Forward Tracking algorithm is performed
without using the drift-time information, reconstructing the hit
positions from the centre of the straw tubes with an uncertainty
equal to their radius. This selects candidates that can be
reconstructed independently of the assumed particle speed and provides
a first estimation of the particle momentum. The drift-time
information is then added and the Forward Tracking is repeated for
different hypotheses on the particle velocity. Explored values of $\beta$
range from 1 to the value corresponding to a $^{4}$He
nucleus with momentum equal to the initial estimate, in steps of 0.05. The latter value is chosen as a compromise
between the highest tracking efficiency and the minimum computing time.
The track candidate is obtained from a fit minimising the \chisq of the spatial residuals of the hits associated with the track and the $\beta$ value corresponding to the minimum fit \chisq is
assigned to the candidate. The quality requirements on the candidates
are simplified by omitting the neural-network-based selection~\cite{Stahl:2017nlr}, which is used in the standard procedure and
optimised for high-momentum particles. 

\begin{figure}
	\centering
    \includegraphics[width=0.495\textwidth]{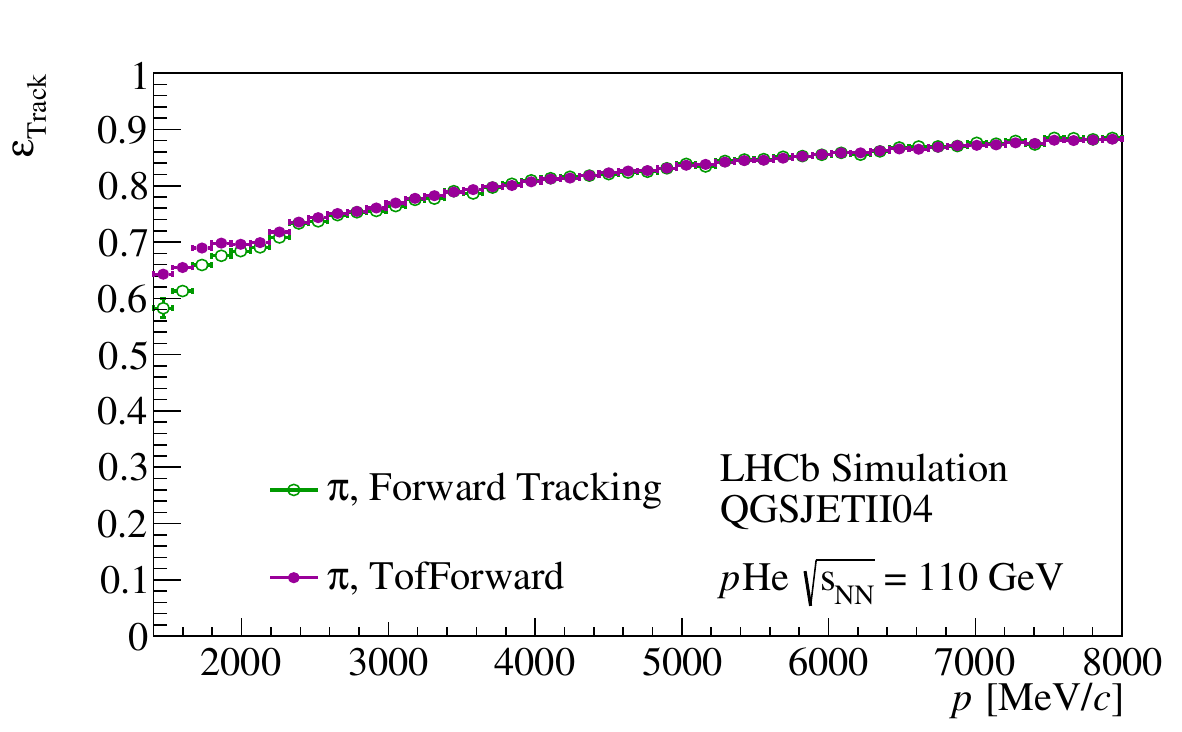}
    \includegraphics[width=0.495\textwidth]{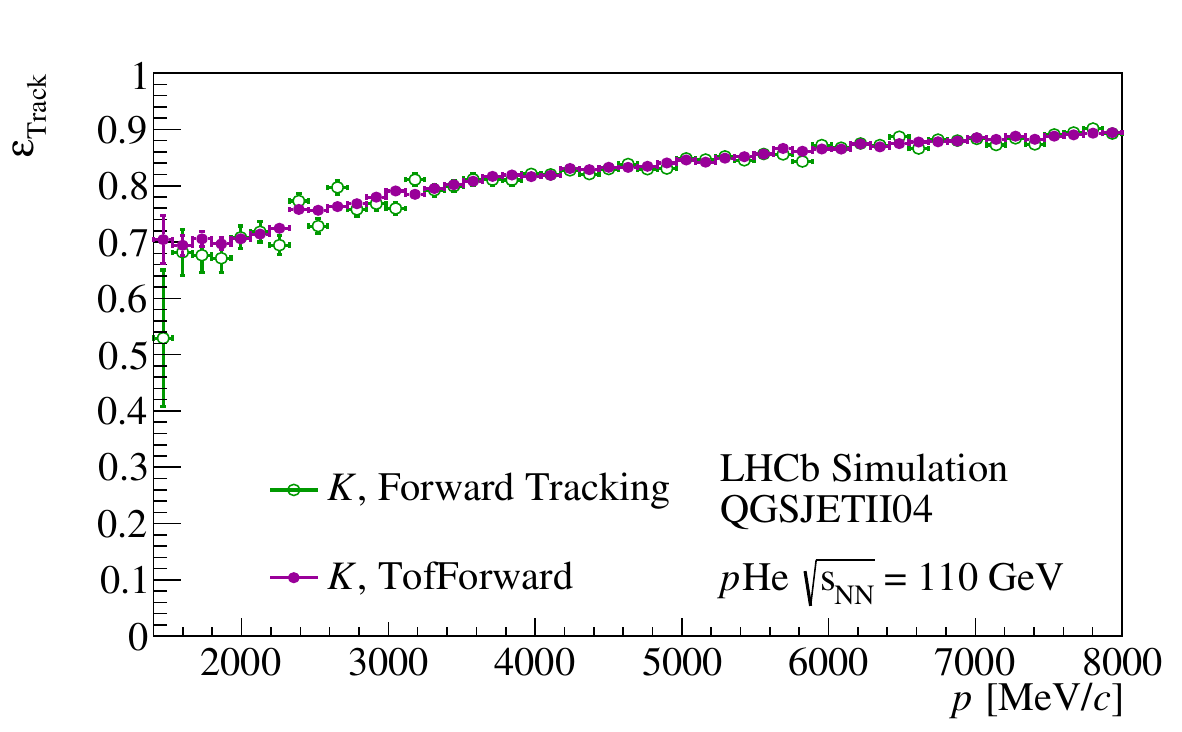}
    \includegraphics[width=0.495\textwidth]{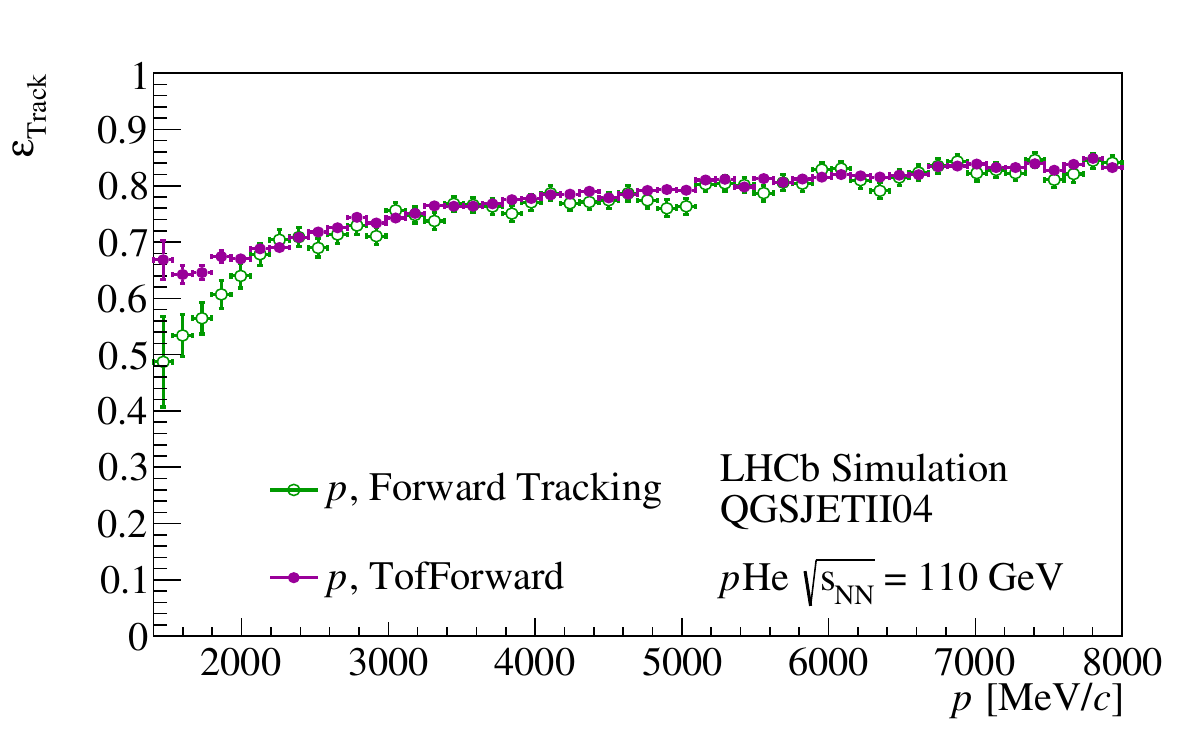}
    \includegraphics[width=0.495\textwidth]{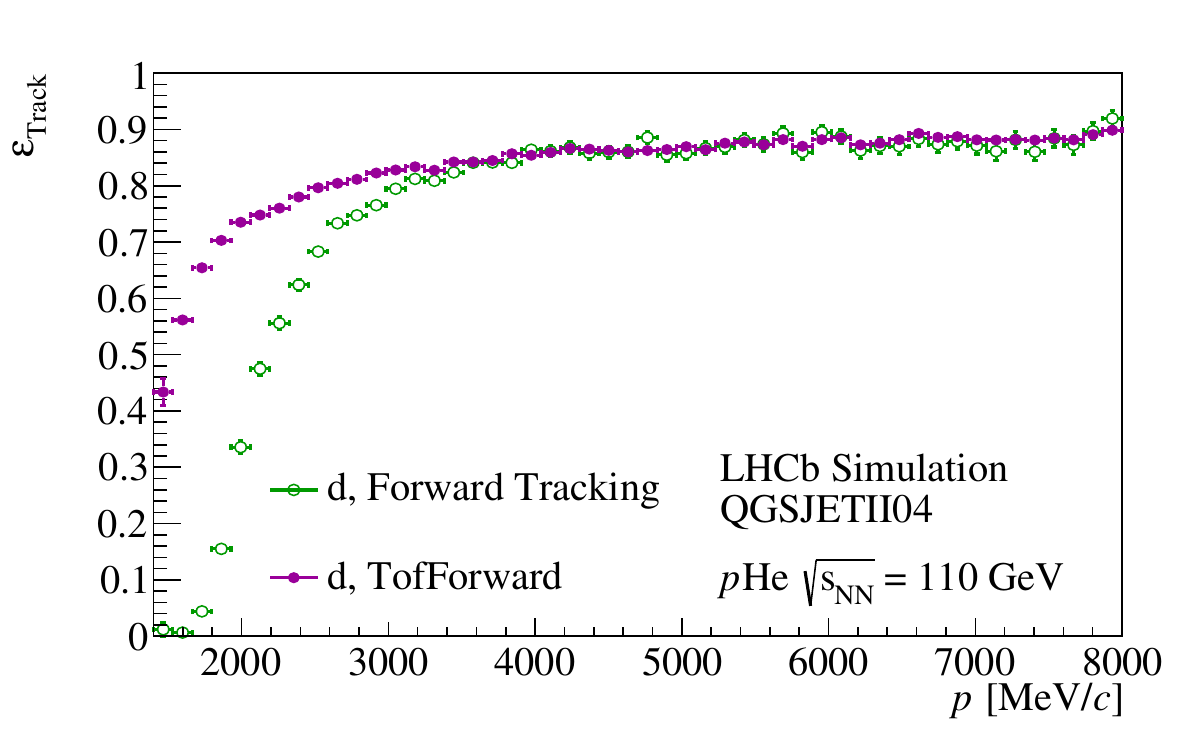}
	\caption{Track reconstruction efficiency as a function of the
          momentum for (top left) pions, (top right) kaons, (bottom
          left) protons and (bottom right) deuterons; the performance
          of (empty markers) the standard Forward Tracking algorithm
          and of  (full markers) the TofForward algorithm is compared.}
	\label{fig:RecoCmp}
\end{figure}

\begin{figure}
	\centering
    \includegraphics[width=0.495\textwidth]{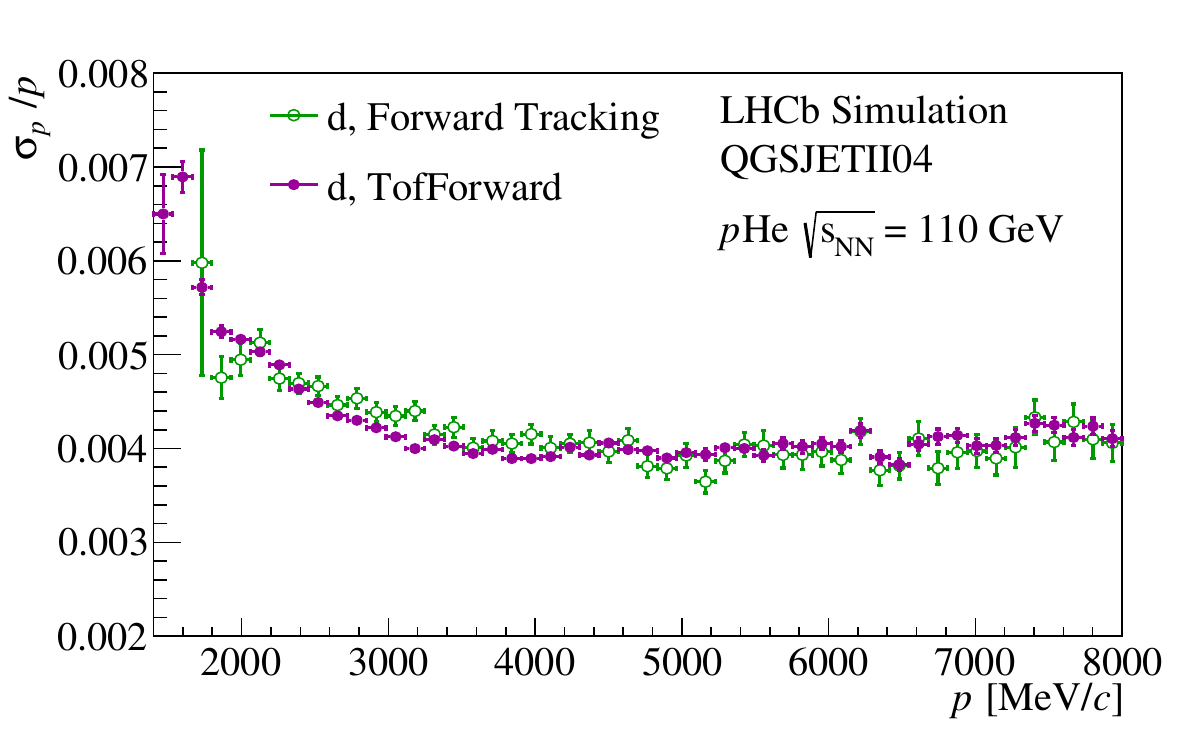}
    \includegraphics[width=0.495\textwidth]{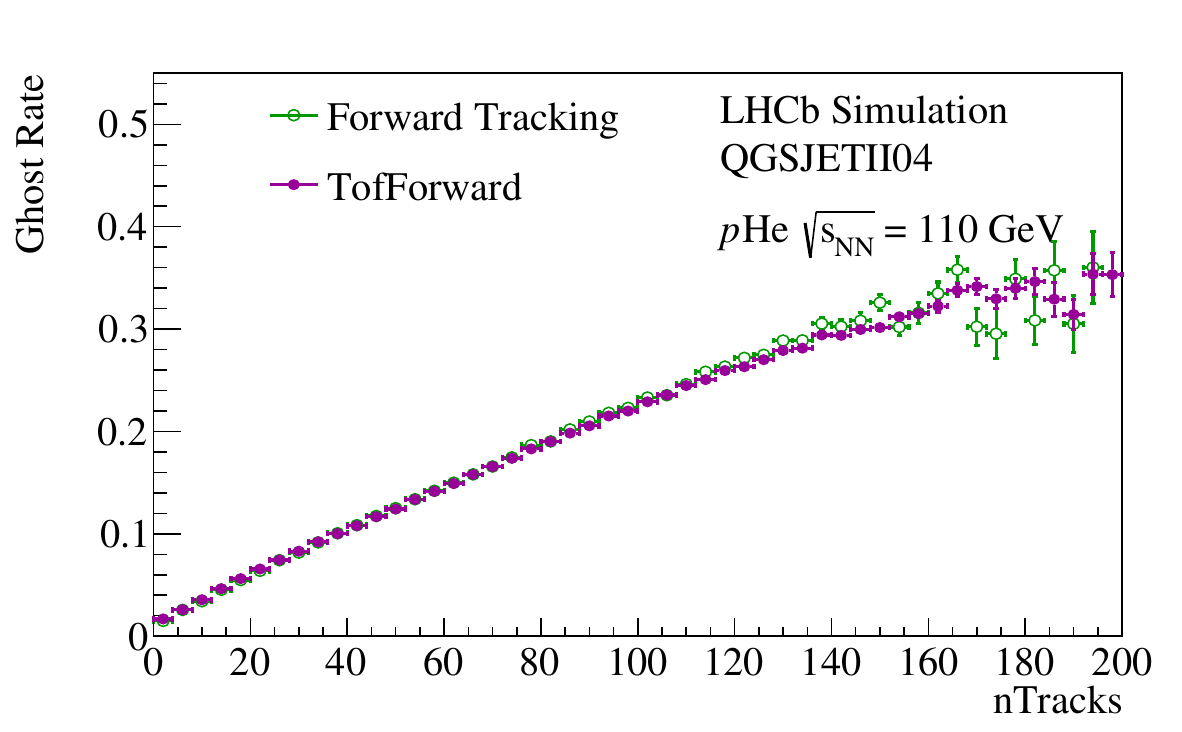}
	\caption{(Left) Momentum resolution as a function of the momentum for
          deuterons in the QGSJET-based simulation
          sample for (empty markers) the standard Forward Tracking algorithm
          and (full markers) the TofForward algorithm. (Right) Average number of ghost tracks
          as a function of the reconstructed track multiplicity is also compared.}
	\label{fig:ResoGhost}
\end{figure}

The tracking efficiency obtained with the TofForward algorithm is compared to the standard reconstruction in Fig.~\ref{fig:RecoCmp}. A significant improvement is achieved for deuterons (protons) with momenta below 4\gevc (2\gevc). The momentum resolution, being dominated at low momentum by multiple scattering, is found to be comparable between the two algorithms, see Fig.~\ref{fig:ResoGhost}. To verify that the looser selection is not affecting the quality of the reconstructed track sample, the rate of so-called ghost tracks, namely candidates built combining hits from different particles, as a function of the event multiplicity, measured as the number of reconstructed tracks (nTracks) in the event, is also compared in Fig.~\ref{fig:ResoGhost}. No significant difference is observed between the two algorithms.

\section{Measurement of the particle speed}
\label{sec:idtool}

In the \lhcb track reconstruction procedure, the track candidates obtained from
the pattern recognition algorithm are refitted using an algorithm
based on a Kalman filter~\cite{LHCb-DP-2019-001} to achieve 
a refined determination of the particle kinematics
by taking into account multiple scattering and energy loss effects.

This algorithm is extended to be able to estimate $\beta$, using
 the simulated sample of standalone deuterons as a testbench.
A given $\beta$ value is tested by running the Kalman filter after
correcting the OT reconstructed drift times for the expected particle 
time of flight along its reconstructed trajectory, assuming
that the particle is produced at the position of closest approach to
the nominal beam line. The $\beta$ estimation is obtained by
minimising the track \chisq of the Kalman filter plus
the normalised residual of the time measurement in the first muon station M1,
\begin{equation}
  \chisq_{\beta}(\beta) = \chisq_{\text{Kalman}}(\beta) + 
\left[ \dfrac{t_{\rm M1}-t_{\rm M1,exp}(\beta)}{\sigma_{\rm M1}} \right]^2\, ,
\end{equation}
where $t_{\rm M1}$ is the observed time in M1 and $t_{\rm M1,exp}$ its
expected value for a given $\beta$ hypothesis;
$\sigma_{\rm M1}=4$~ns is the M1 time resolution.

Track fits with different values of $\beta$ are performed with a fixed number of detector
hits, starting from those associated to the track by the TofForward
algorithm. The drift-time information is always used, even when the
reconstructed value is apparently above its physically allowed maximum value of 35\ns. These two prescriptions are needed to obtain a well-behaved $\beta$
dependence of $\chisq_{\beta}$, with a profile that
approximates a parabola in the region surrounding the minimum.
To speed up the algorithm, the
minimum of $\chisq_{\beta}$ is found with a gradient descent
algorithm, starting from $\beta=1$ and a negative step of 0.01.
The sampled $\beta$ values are limited between 1.1 and the value 
corresponding to a mass of 4\gevcc for the measured particle momentum.  
The values around the minimum are
fitted with a second-order polynomial to obtain the estimated value 
$\hat{\beta}$ and its uncertainty $\sigma(\hat{\beta})$.
To reduce biases from hits that are incorrectly assigned to the track, the
$\hat{\beta}$ determination is updated if some hits in the track fit for $\beta=\hat{\beta}$ exhibit a residual larger than three
standard deviations. The procedure is repeated up to four times, removing
up to two outlier hits on every iteration.

\begin{figure}[tb]
	\centering 
        \includegraphics[width=0.49\textwidth]{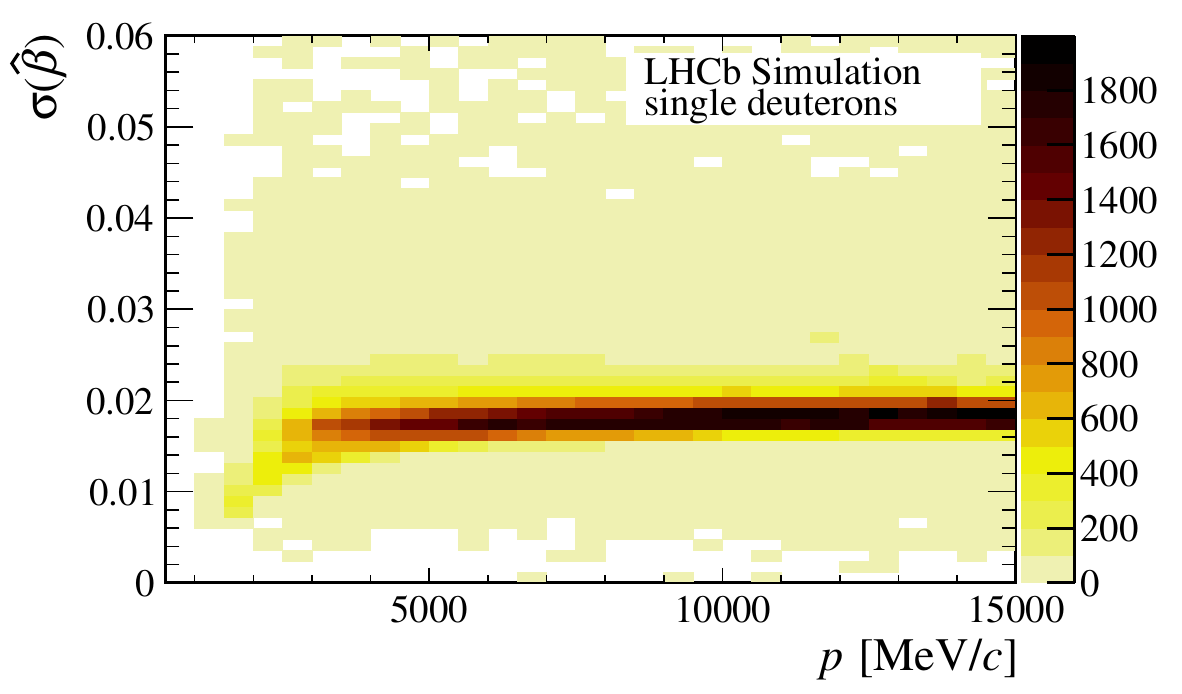}
        \includegraphics[width=0.49\textwidth]{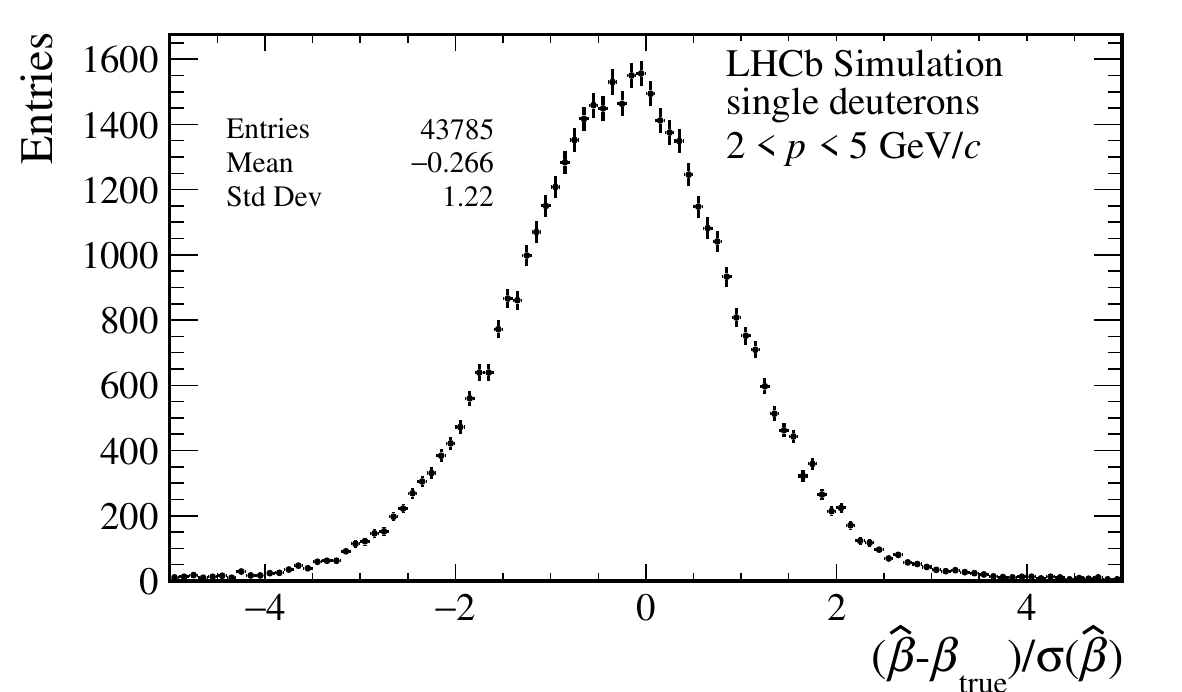}
	\caption{ Performance of the $\beta$ determination on the
          simulated sample of single
          deuteron tracks: (left) estimated
          uncertainty   $\sigma(\hat{\beta})$ as a function of
          momentum; (right) normalised residuals for
          low-momentum particles.}
	\label{fig:betaSingleDeuterons}
\end{figure}

\begin{figure}[tb]
	\centering
    \includegraphics[width=\textwidth]{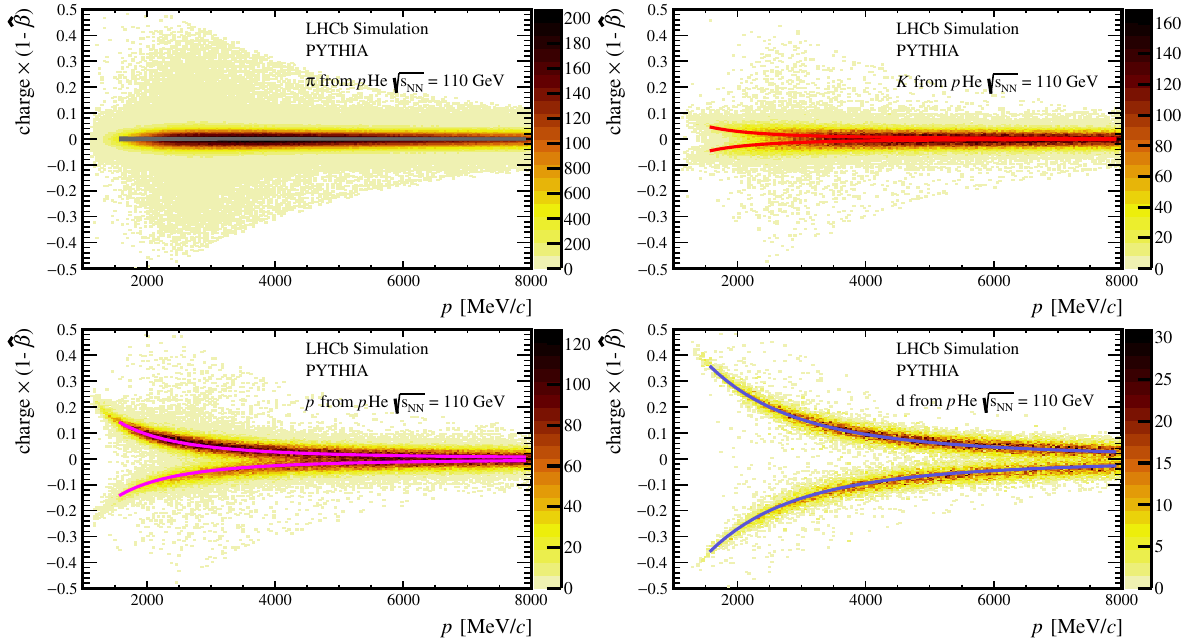}
	\caption{ Difference with respect to 1 of $\hat{\beta}$, multiplied by the particle charge, as
a function of the reconstructed momentum for four hadron species in
the $\pythia$-based simulated sample:  (top left) pions,
(top right) kaons, (bottom left) protons and  (bottom right)
deuterons. The overlaid curves show the expected relations.}
	\label{fig:betaMC}
\end{figure}

Figure~\ref{fig:betaSingleDeuterons} shows the performance for simulated
single deuterons: $\sigma(\hat{\beta})$ is typically
below 0.02 for $p<3 \gevc$ ($\beta<0.73$ ) and the bias on the $\beta$ determination, defined as the difference between the estimated value $\hat{\beta}$ and the true value $\beta_{\text{true}}$, is well below its average uncertainty. Results for the $\beta$
measurements on  different hadron species in the \pythia-based simulation are shown in  
Fig.~\ref{fig:betaMC}. 


\begin{figure}[tb]
	\centering 
        \includegraphics[width=.97\textwidth]{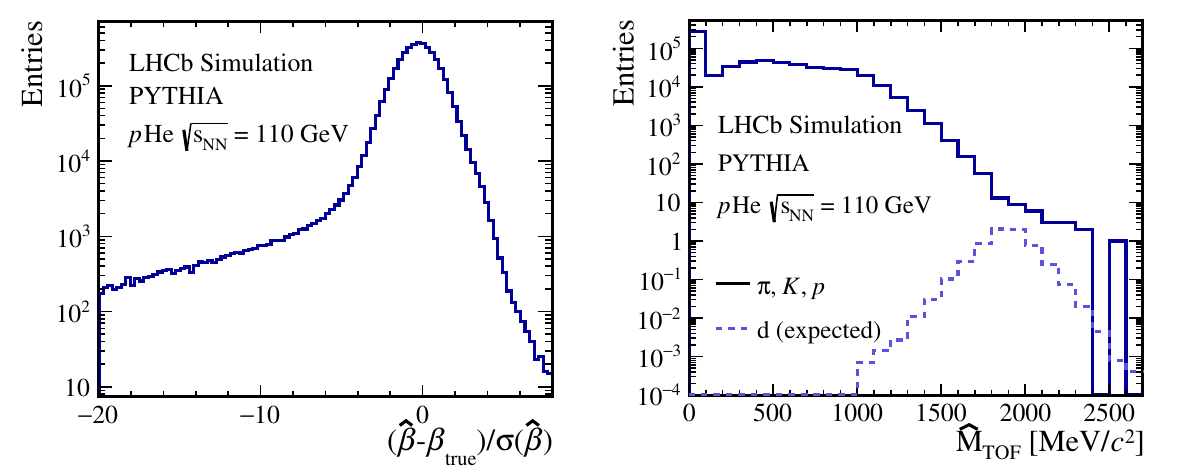}
	\caption{ (Left) Residual of the $\beta$ estimation
        normalised by its estimated uncertainty for simulated
        positively charged particles with momentum below 4$\gevc$. (Right) 
        Distribution of the corresponding  estimated mass \ensuremath{\hat{\rm M}_{\text{TOF}}}\xspace, separately for
        deuterons and lighter particles, assuming that a deuteron is produced approximately one out of every $10^4$ collisions.  }
	\label{fig:betaTail}
\end{figure}

While the simulated collision samples are
enriched with deuterons, the antideuteron/antiproton production ratio
for  $p<5 \gevc$  in fixed-target collisions at the \lhcb experiment is expected to be
between $10^{-4}$ and  $10^{-3}$~\cite{PhDLucarelli}. The isolation of a deuteron sample
requires more discriminating power against lighter hadrons than the one
achievable with the $\hat{\beta}$ estimator alone.
This is illustrated in Fig.~\ref{fig:betaTail}, showing the distribution
of the estimated particle mass from the time-of-flight and momentum
measurements
\begin{equation}
\hatM \equiv \dfrac{p}{c} ~ \sqrt{\dfrac{1}{\min(\hatbeta,1)^2}-1}\, ,
\label{eq:mtof}
\end{equation}
for the expected deuteron signal and for lighter particles. 
A significant background to deuteron identification is expected
from cases where $\beta$ is underestimated by an amount
far beyond its estimated uncertainty.

\begin{figure}[b]
	\centering
    \begin{subfigure}{0.49\textwidth}
    \captionsetup{labelformat=empty, skip=-15pt}
    \includegraphics[width=\textwidth]{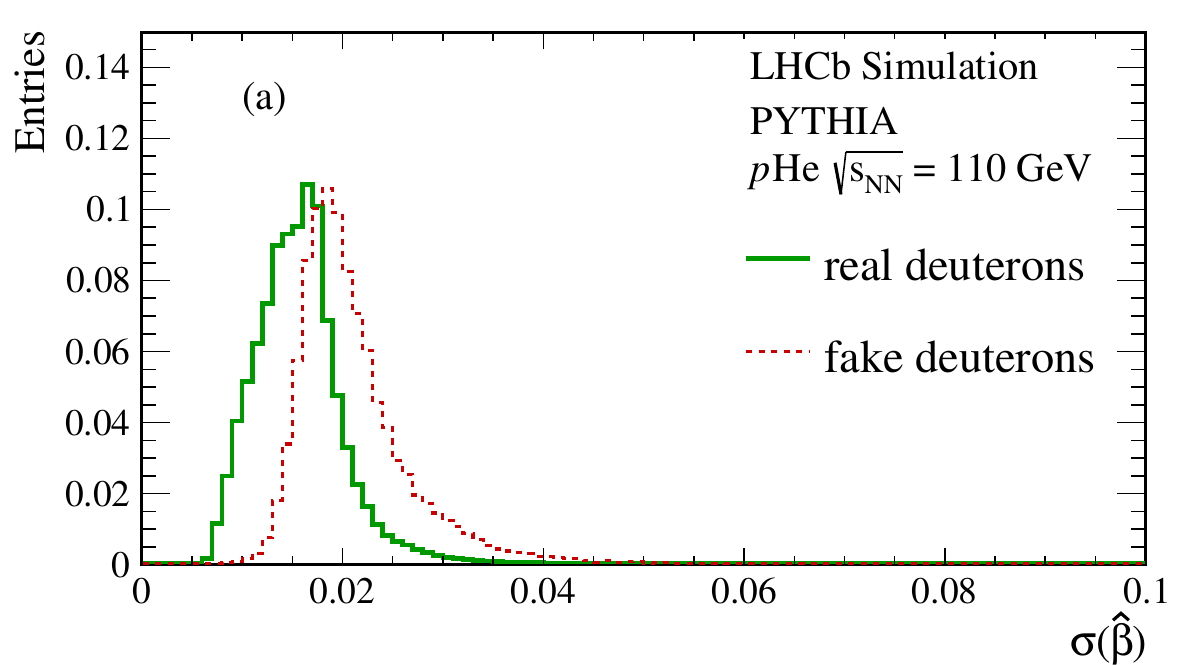}
    \caption{\hspace{-1em}} \label{fig:betaerr}
    \end{subfigure}
    \begin{subfigure}{0.49\textwidth}
    \captionsetup{labelformat=empty, skip=-15pt}
    \includegraphics[width=\textwidth]{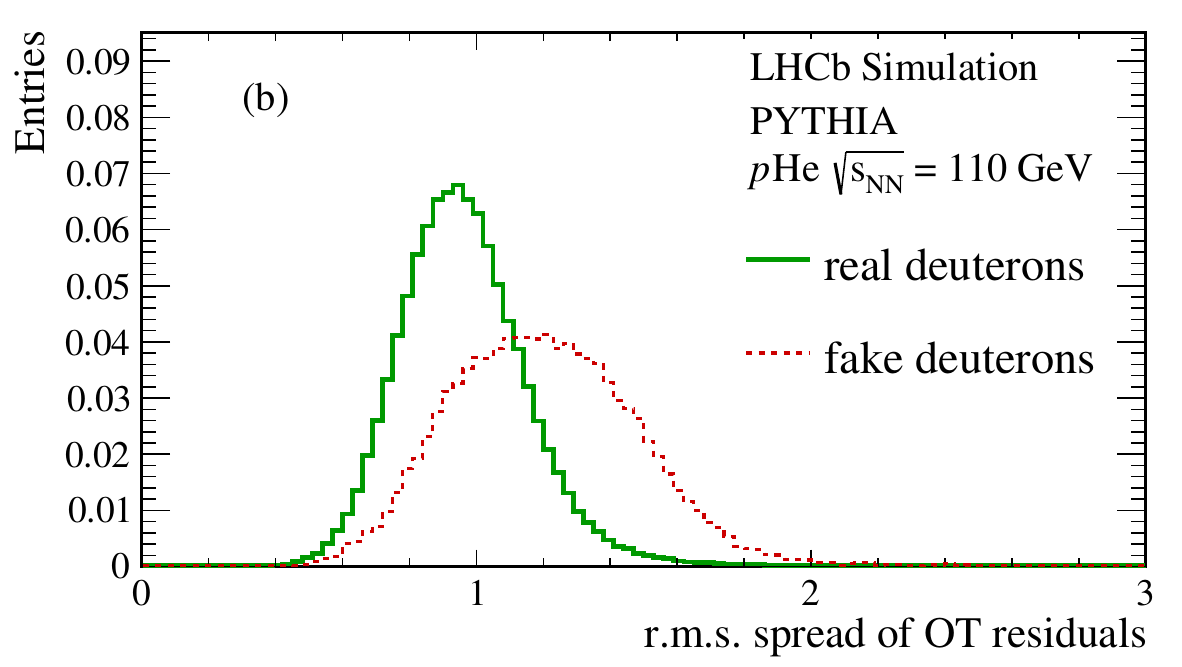}
    \caption{\hspace{-1em}} \label{fig:OTresSpr}
    \end{subfigure}
    \begin{subfigure}{0.49\textwidth}
    \captionsetup{labelformat=empty, skip=-15pt}
    \includegraphics[width=\textwidth]{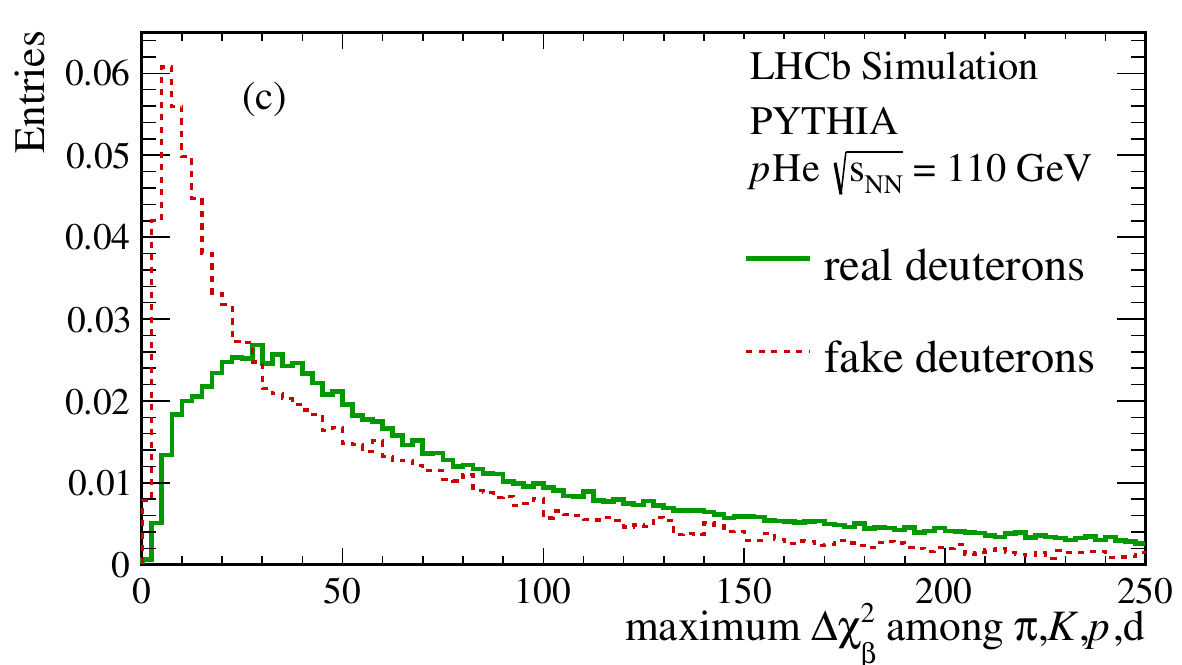}
    \caption{\hspace{-1em}} \label{fig:maxdchi2}
    \end{subfigure}
    \begin{subfigure}{0.49\textwidth}
    \captionsetup{labelformat=empty, skip=-15pt}
    \includegraphics[width=\textwidth]{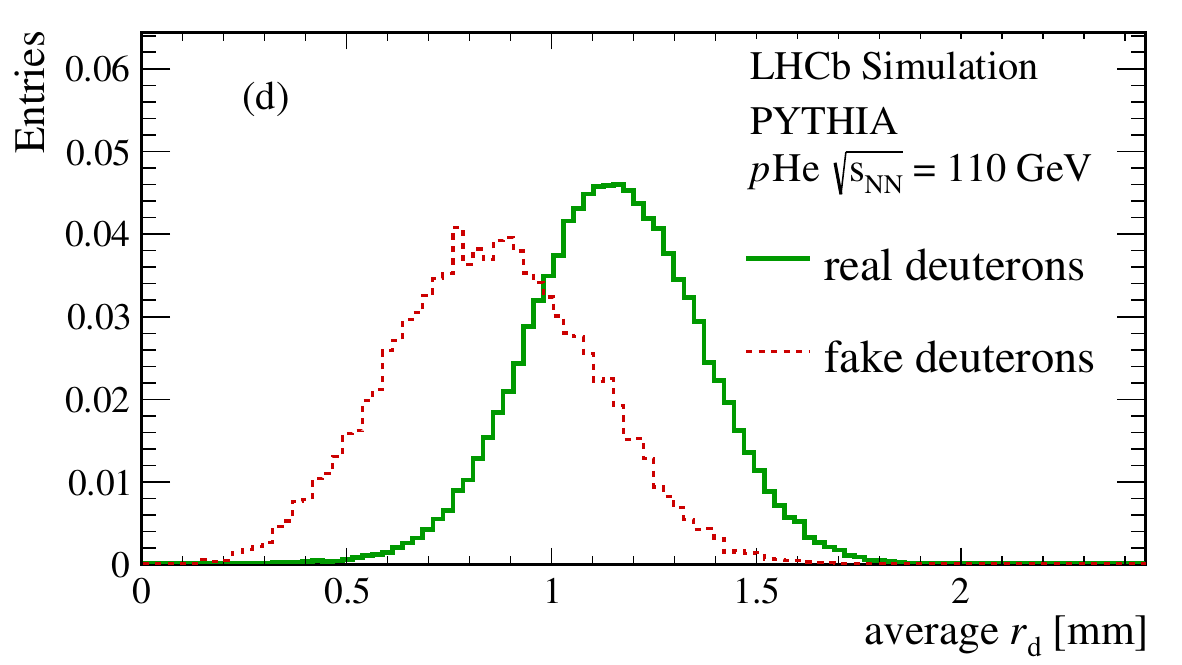}
    \caption{\hspace{-1em}} \label{fig:OTradAv}
    \end{subfigure}
    \begin{subfigure}{0.49\textwidth}
    \captionsetup{labelformat=empty, skip=-15pt}
    \includegraphics[width=\textwidth]{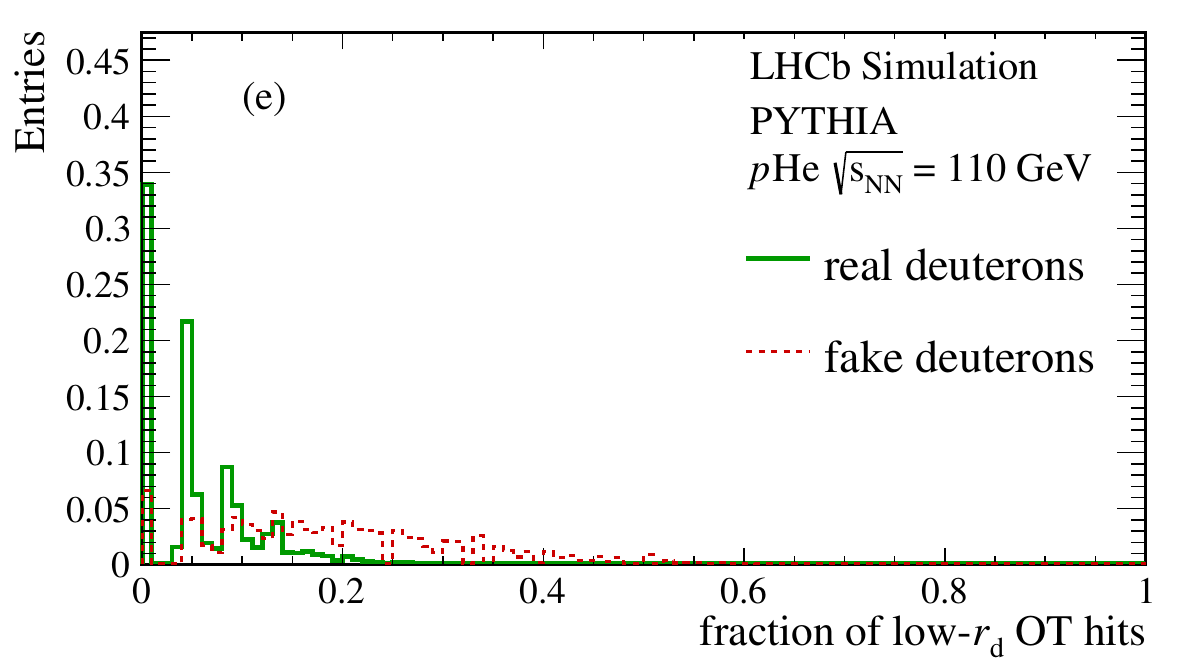}
    \caption{\hspace{-1em}} \label{fig:OTradLow}
    \end{subfigure}
    \begin{subfigure}{0.49\textwidth}
    \captionsetup{labelformat=empty, skip=-15pt}
    \includegraphics[width=\textwidth]{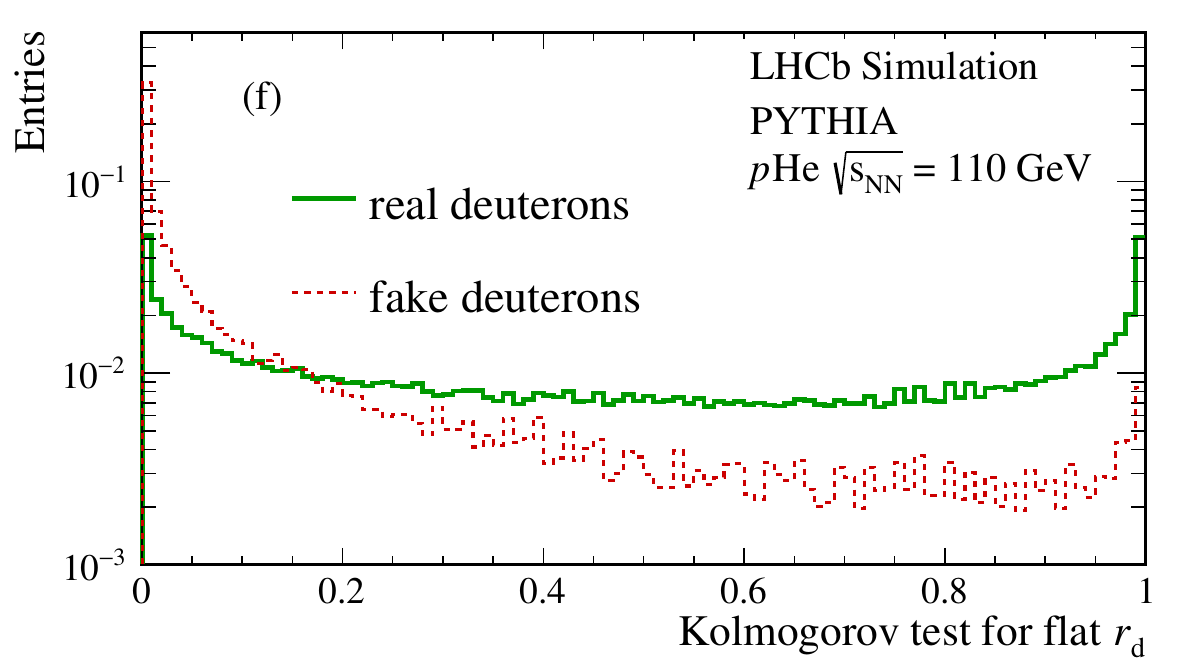}
    \caption{\hspace{-1em}} \label{fig:OTradKolm}
    \end{subfigure}
	\caption{ Distributions in simulation of some of the observables used to
          discriminate genuine deuterons from lighter particles with an \ensuremath{\hat{\rm M}_{\text{TOF}}}\xspace
          value compatible with the deuteron mass (fake deuterons), for tracks with
          momentum lower than 3$\gevc$.}
	\label{fig:fakeD}
\end{figure}

This background is studied in simulation
by selecting those hadrons among pions, kaons, and protons whose \hatM
value is compatible with the deuteron
mass within 400\mevcc, called fake deuterons in the following. 
Several observables related to the quality of the track fit and $\beta$ determination 
are considered in order to identify fake deuterons and improve the $\beta$ estimation.
Observables that are found to provide significant discrimination
between fake and real  deuterons of equal momentum include the uncertainty $\sigma(\hat{\beta})$, shown in Fig.~\ref{fig:betaerr}, the number of OT hits, the quality of the reconstructed
track (track \chisq, average and spread of residuals in the OT as in Fig.~\ref{fig:OTresSpr}, variation of the track \chisq
and of the number of outliers between the track fits with $\beta=1$ and $\beta=\hat{\beta}$), the maximum difference
in $\chisq_{\beta}$ among four hadron hypotheses ($\pi$, $K$, $p$, $d$) in Fig.~\ref{fig:maxdchi2},
and the distribution of the reconstructed drift radius $r_{\rm d}$  in the OT straw
tubes, namely the distance from the straw tube wire
corresponding to the reconstructed drift time.
The $r_{\rm d}$ observable is particularly important since for light hadrons with
underestimated $\beta$, the reconstructed drift times are
underestimated, while a flat distribution of the drift radius between
0 and its maximum value of 2.5~mm is expected for properly reconstructed tracks.
This is exploited by computing four observables from the OT hits
associated to each track candidate: the average
and r.m.s. spread of drift radii in Fig.~\ref{fig:OTradAv}, the fraction of $r_{\rm d}$ values lower than $0.2\mm$ in the distribution of Fig.~\ref{fig:OTradLow} and
the result of a Kolmogorov--Smirnov test~\cite{Smirnov} of the  $r_{\rm d}$ values  against a
uniform distribution between 0 and 2.5~mm in Fig.~\ref{fig:OTradKolm}.

\begin{figure}[tb]
	\centering
        \includegraphics[width=.6\textwidth]{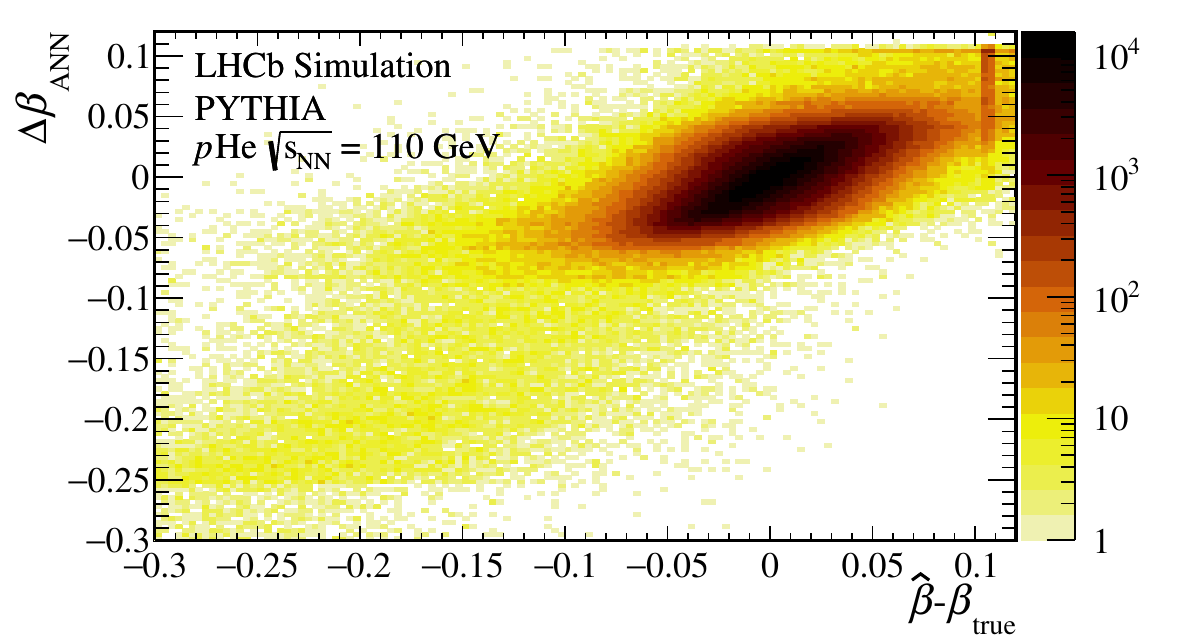}
	\caption{ Performance of the \ensuremath{\Delta\beta_{\text{ANN}}}\xspace predictor on
          simulated collisions. The response is shown for all track
          candidates with momentum lower than 5$\gevc$. The structure at $\Delta\beta=0.1$ 
          is an artifact from the limit $\beta<1.1$ on the sampled values.} 
	\label{fig:nnValidMC}
\end{figure}

A total of 13 variables, including the particle momentum, are fed
into an artificial neural network (ANN), trained on simulated collisions to
predict the residual  $\dbeta=\hat{\beta}-\beta_{\text{true}}$, where $\beta_{\text{true}}$ is
the true value from simulation. To avoid that any requirement on the
network output could bias the particle identification by producing 
fake peaks in the \hatM distribution, no input is
provided about the particle type and its true or measured speed.  
The goal is to obtain an assessment of the accuracy of the $\beta$
estimation from observables related to the quality of the fit
without prior knowledge of the estimated $\beta$ value.

\begin{figure}[tb]
	\centering
        \includegraphics[width=0.95\textwidth]{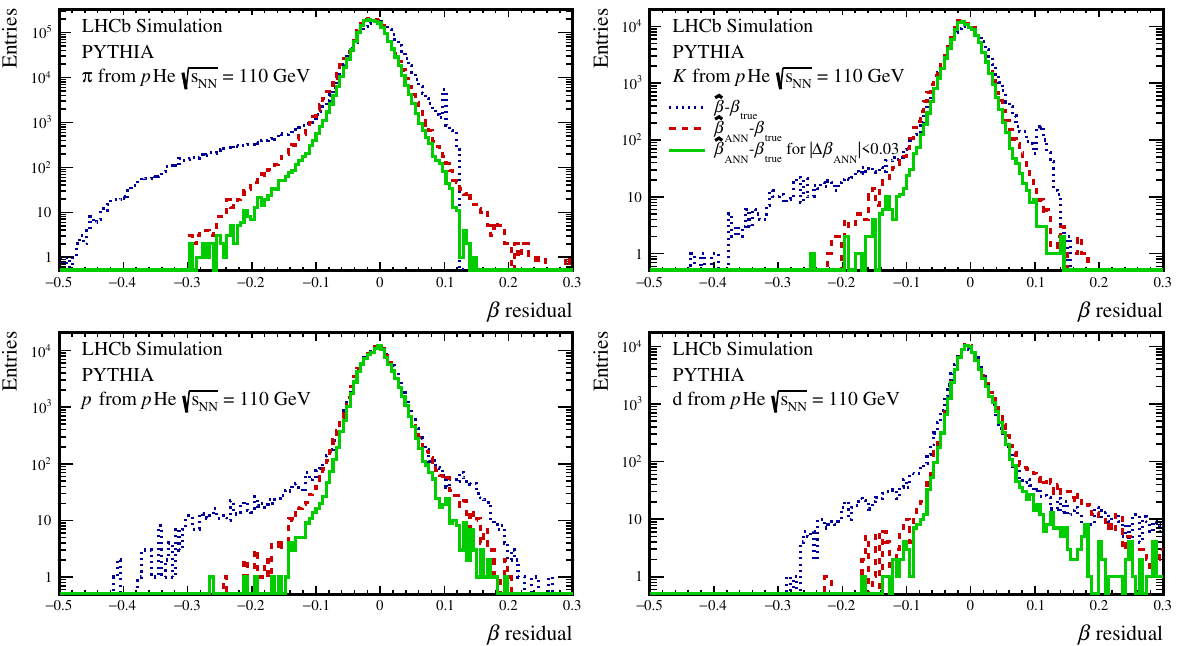}
	\caption{ Impact of the use of the \ensuremath{\Delta\beta_{\text{ANN}}}\xspace predictor on the accuracy of 
          the $\beta$ determination for different hadron species in
          simulated collisions, for tracks with momentum lower than 5$\gevc$.}
	\label{fig:nndbetaMC}
\end{figure}

The network, implemented using 
\textsc{Keras}~\cite{Keras} with \tensorflow~\cite{TF} backend, consists 
of two tanh-activated hidden layers with 32 neurons each and is trained on 60\% of the simulated collision sample. 
The performance of the resulting predictor \dbetaNN is illustrated in
Fig.~\ref{fig:nnValidMC} in the remaining 40\% of simulated collisions.
The results show that the predictor identifies most of the cases
where $\beta$ is badly underestimated, so that fake deuterons can be
suppressed by requiring a small value of $|\dbetaNN|$.
Moreover, it shows a good correlation with the actual residual even in
the bulk of the distribution, allowing the $\beta$ estimator to be refined as 
\begin{equation}
  \hatbetaNN \equiv \hatbeta - \dbetaNN.
\end{equation}
The resulting improvement of the $\beta$ estimation  is illustrated in  Fig.~\ref{fig:nndbetaMC}.

\section{Performance}
\label{sec:performance}

The performance obtained in simulation is verified  on a sample of 
\pHe collisions at \mbox{\sqsnn=110\gev}, collected in 2016
with loose online selection requirements. The sample consists of about 700 million
events, corresponding to an integrated luminosity of 0.5\invnb~\cite{LHCb-PAPER-2018-031}.

\begin{figure}[tb]
  \centering
  \includegraphics[width=\textwidth]{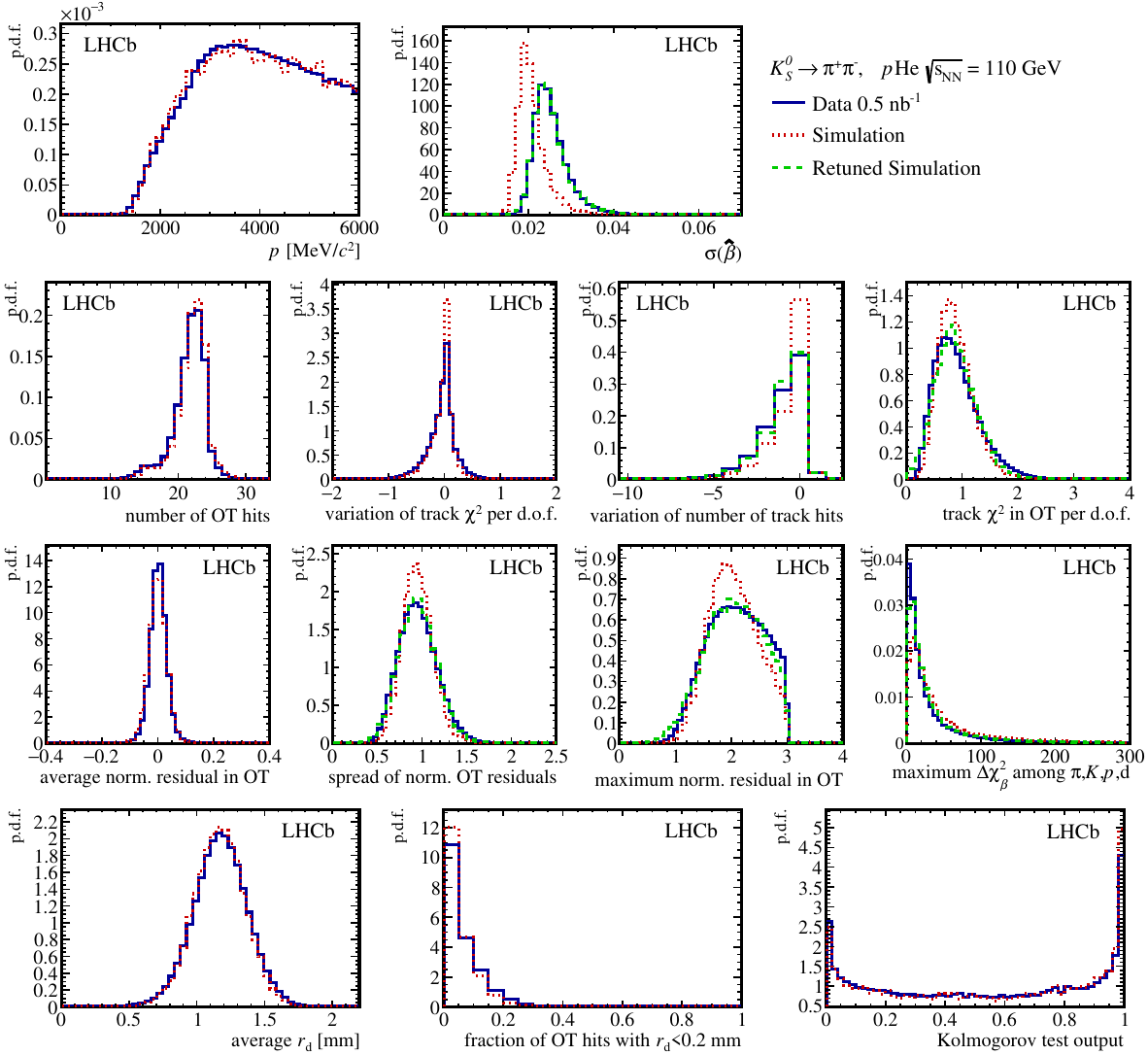}
  \caption{
    Distributions of the 13 observables, described in Sec.~\ref{sec:idtool}, 
    used as input to the \ensuremath{\Delta\beta_{\text{ANN}}}\xspace predictor network.
    Data and simulation are compared
    for candidate pion tracks from $\KS$ decays
    with momentum below 6$\gevc$. For six variables, notably
    $\sigma(\hat{\beta})$, an empirical correction is applied to the
    simulated observables to improve the agreement with data.} 
  \label{fig:KSdataMC}
  \end{figure}

The $\beta$ determination procedure is first verified on 
identified pion and proton tracks from $\decay{\KS}{\pip \pim}$ and
\decay{\Lz}{\proton\pim} decays, for which the identity of the particles can be determined from the decay 
kinematics~\cite{Podolanski01011954}.
Candidates are selected based on their 
incompatibility with originating from the primary collision vertex, the  
reconstructed mass of the decaying resonance, and the quality of the
reconstructed decay vertex. The purity of these calibration samples
exceeds 98\%. The same exercise is performed on the simulated sample
and the distributions of the observables used in
this study are shown in Fig.~\ref{fig:KSdataMC} to be in a good overall agreement with data.
The largest difference that can be noticed is a larger average
value of $\sigma(\hat{\beta})$ in data. Comparing the $\hat{\beta}$
estimation with  the expected $\beta$ from the measured
momentum and the known mass of the identified pion or proton, a
negative bias is seen,  of order 0.01 and larger than the one observed in simulation.
Both effects, the differences in $\hat{\beta}$ and $\sigma(\hat{\beta})$ values, are 
compatible with a residual miscalibration of the OT
response of $\order(0.1)\ns$.  The bias on $\hat{\beta}$ is corrected 
by parametrising it with an empirical function of $p$ and
$\hat{\beta}$ tuned on the calibration sample.

\begin{figure}[tb]
  \centering
  \includegraphics[width=0.6\textwidth]{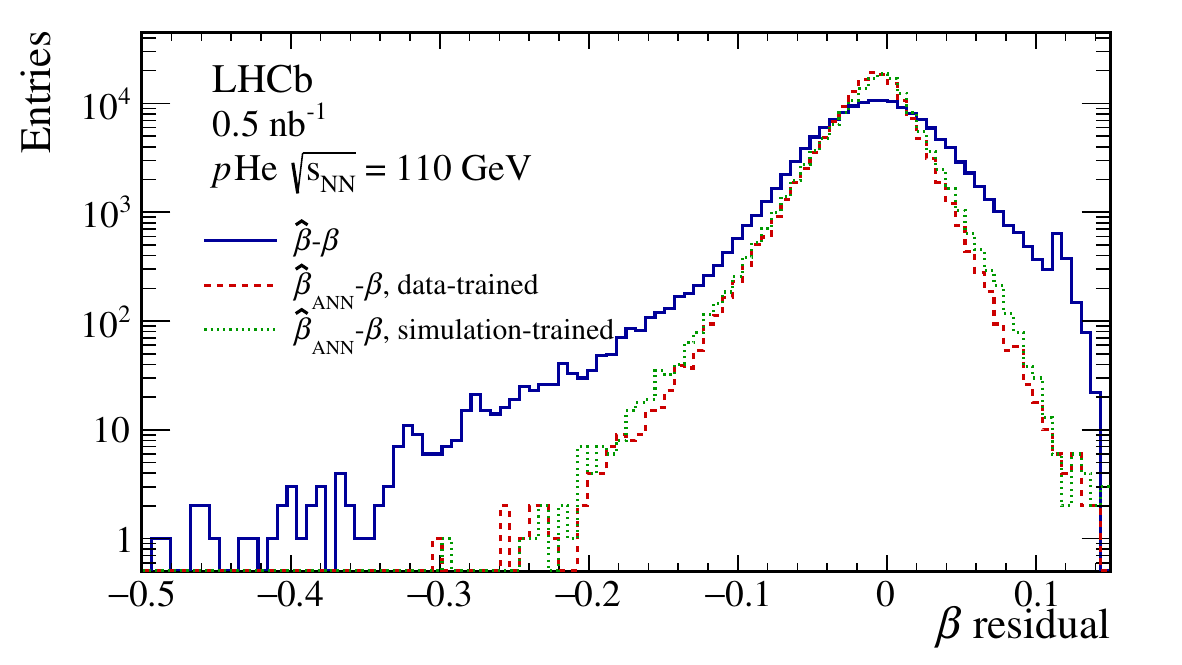}
  \caption{ 
    Performance of the \ensuremath{\Delta\beta_{\text{ANN}}}\xspace predictor on the calibration sample,
    for momenta lower than 5$\gevc$. The plot shows the difference 
    between the $\beta$ estimation and the expected value
    $\beta=p/\sqrt{\left[p^2+(Mc)^2\right]}$, obtained from the measured momentum and
    the known mass $M$ of the identified pion or proton.  
    The comparison is performed on the 20\% of calibration data not
    used for the training. 
}
  \label{fig:nnDATAMC}
\end{figure}
  
Two \dbetaNN predictor networks are retrained making use of the calibration data.
The first is directly trained on the calibration tracks, using
80\% of the sample.
The second network is trained on the simulated collisions after correcting for
the observed differences in the distribution of $\sigma(\hat{\beta})$
and other five input variables, as shown in  Fig.~\ref{fig:KSdataMC}. 
The  $\sigma(\hat{\beta})$ value is scaled by a factor
$(1+k\hat{\beta}^2)$, where $k=0.22$ is obtained from the calibration
pion sample and the $\beta$ dependence from a miscalibration effect is
assumed. For the other variables, the correction consists of adding a Gaussian-smeared value to
reproduce the average and r.m.s. spread of the distribution observed in the calibration data.

Figure~\ref{fig:nnDATAMC} presents a comparison of the performance of the two networks 
on the 20\% of the calibration data that was not used during training.
Though the data-driven predictor has
slightly better performance, the network trained on simulation identifies
correctly most of the fake deuterons in the calibration sample
and is preferred in the following to avoid any overtraining in data. Moreover, its training includes deuterons, which are the main goal of
this study and for which no data calibration sample is available.

\begin{figure}[tb]
	\centering
        \includegraphics[width=\textwidth]{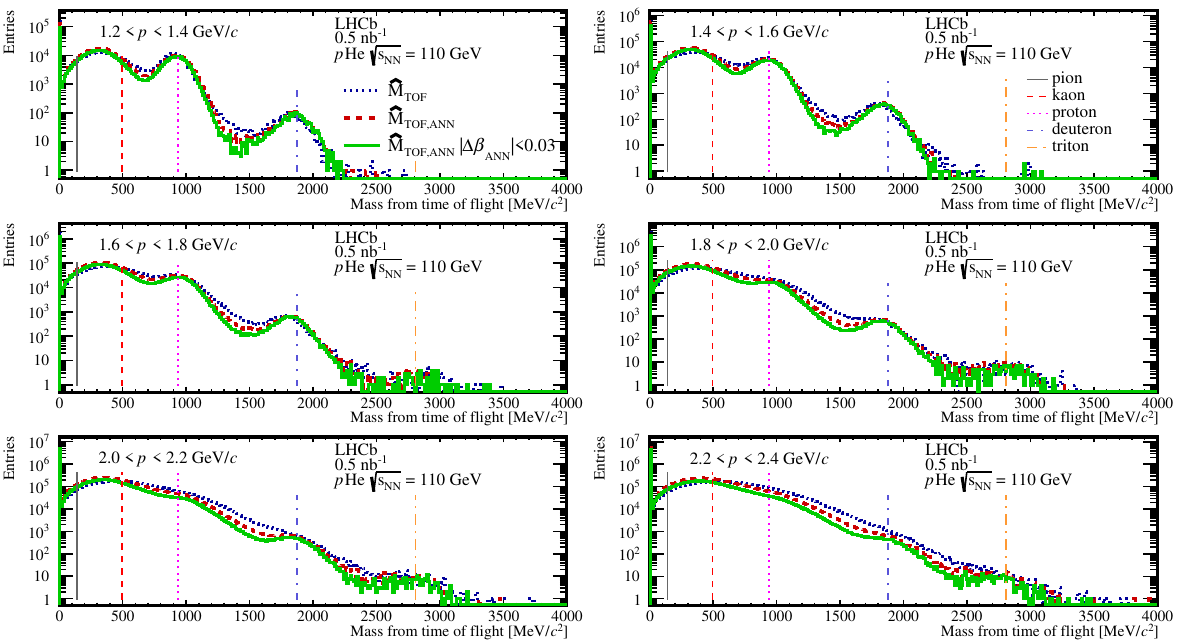}
	\caption{Estimated particle mass from the time of flight for 
          reconstructed tracks of positively charged particles in the \ensuremath{\proton\text{He}}\xspace collision
          sample, for different momentum intervals. \ensuremath{\hat{\rm M}_{\text{TOF,ANN}}}\xspace is
          obtained by replacing \ensuremath{\hat{\beta}}\xspace with 
          \ensuremath{\hat{\beta}_{\text{ANN}}}\xspace in Eq.~\ref{eq:mtof}. 
          The vertical lines correspond to the known mass of pions,
          kaons, protons, deuterons and tritons.}
	\label{fig:tofMassDATA}
\end{figure}

\begin{figure}[tb]
	\centering
        \includegraphics[width=0.6\textwidth]{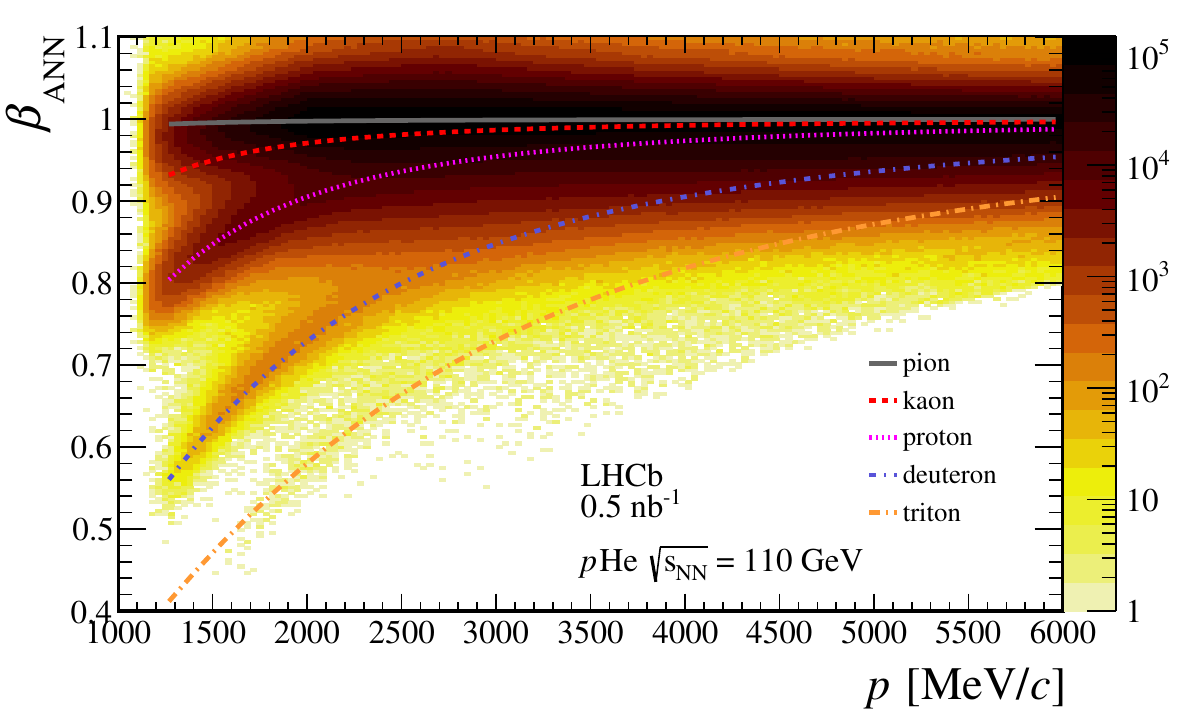}
	\caption{Estimated speed versus momentum for 
          reconstructed tracks of positively charged particles in the \ensuremath{\proton\text{He}}\xspace collision
          sample, using the \ensuremath{\hat{\beta}_{\text{ANN}}}\xspace estimator and requiring
          $|\ensuremath{\Delta\beta_{\text{ANN}}}\xspace|<0.03$. The overlaid curves show the expected
          relations for different hadron species. Proton,
          deuteron and triton production can be seen.
          The momentum-dependent cutoff at low $\beta$ 
          corresponds to the maximum mass hypothesis
          allowed in the algorithm.}
	\label{fig:tofDATA}
\end{figure}

The $\beta$ determination is finally performed on the whole sample of reconstructed 
low-momentum tracks ($p<6\gevc$) with loose preselection requirements (at least 6 OT hits
and $\sigma(\hat{\beta})<0.1$).
Figure~\ref{fig:tofMassDATA} shows the reconstructed mass values in
different momentum intervals for positively charged particles, using the $\hat{\beta}$
or \hatbetaNN estimators and with or without an additional
requirement on the \dbetaNN predictor.
A clear deuteron signal is visible for $p<2.5\gevc$, and an excess of tracks in the
region of the triton mass is also observed. These ion candidates 
are expected to be mostly produced by the fragmentation of the helium target and are 
clearly visible only in the sample with reconstructed tracks of positively charged particles.

As expected from the studies on simulation, the neural-network-based
predictor \hatbetaNN provides better 
discrimination among particle species than  \hatbeta, and the 
deuteron identification is further improved by 
the \dbetaNN requirement. 
The distribution of \hatbetaNN as a function of momentum for
tracks of positively charged particles with $|\dbetaNN|<0.03$
is shown in Fig.~\ref{fig:tofDATA}.

A few He ion candidates are also identified in the \pHe sample, thanks to their ionisation losses in the 
Si-based tracking detectors~\cite{LHCb-DP-2023-002}. The OT timing response for particles with atomic number $Z=2$ 
(or larger) is biased toward shorter drift times, due to their higher ionisation with respect to $Z=1$ particles. 
In the explored low-momentum region, the biased reconstructed speed roughly corresponds
to that expected  for protons. Therefore, the time-of-flight technique is not suitable by itself to 
discriminate low-momentum He nuclei from lighter particles, though it could provide some discrimination between $^{3}$He and  $^{4}$He candidates. For He nuclei with momentum larger than 10\gevc, the measured time of flight in the OT is apparently shorter
than that expected for $\beta=1$, a feature that is already exploited in the He identification tool described in Ref.~\cite{LHCb-DP-2023-002}.

\section{Conclusions}
\label{sec:conclusions}

In this paper, a newly developed identification tool for light ions is
presented, exploiting the available timing information of the \lhcb
detector operated in the LHC Run~2 to distinguish low-momentum
deuterons from lighter hadrons by their time of flight. 
The tool performance is demonstrated by showing its capability to
identify a signal for deuteron and triton production in  
\pHe collisions at \mbox{\sqsnn=110~\gev}.

The benchmark sample has been chosen as it reproduces collisions of cosmic rays on the
interstellar medium at the energy scale corresponding to the capabilities of the AMS-02 spectrometer.
Though this time-of-flight technique in the \lhcb detector can be exploited only in a limited
kinematic region, it opens the way to unique and much awaited measurements of light antinuclei
production at this energy scale, constraining the expected
flux of secondary cosmic antideuterons.

\section*{Acknowledgements}
%
%
\noindent We express our gratitude to our colleagues in the CERN
accelerator departments for the excellent performance of the LHC. We
thank the technical and administrative staff at the LHCb
institutes.
We acknowledge support from CERN and from the national agencies:
ARC (Australia);
CAPES, CNPq, FAPERJ and FINEP (Brazil); 
MOST and NSFC (China); 
CNRS/IN2P3 (France); 
BMBF, DFG and MPG (Germany); 
INFN (Italy); 
NWO (Netherlands); 
MNiSW and NCN (Poland); 
MCID/IFA (Romania); 
MICIU and AEI (Spain);
SNSF and SER (Switzerland); 
NASU (Ukraine); 
STFC (United Kingdom); 
DOE NP and NSF (USA).
We acknowledge the computing resources that are provided by ARDC (Australia), 
CBPF (Brazil),
CERN, 
IHEP and LZU (China),
IN2P3 (France), 
KIT and DESY (Germany), 
INFN (Italy), 
SURF (Netherlands),
Polish WLCG (Poland),
IFIN-HH (Romania), 
PIC (Spain), CSCS (Switzerland), 
and GridPP (United Kingdom).
We are indebted to the communities behind the multiple open-source
software packages on which we depend.
Individual groups or members have received support from
Key Research Program of Frontier Sciences of CAS, CAS PIFI, CAS CCEPP, 
Fundamental Research Funds for the Central Universities,  and Sci.\ \& Tech.\ Program of Guangzhou (China);
Minciencias (Colombia);
EPLANET, Marie Sk\l{}odowska-Curie Actions, ERC and NextGenerationEU (European Union);
A*MIDEX, ANR, IPhU and Labex P2IO, and R\'{e}gion Auvergne-Rh\^{o}ne-Alpes (France);
Alexander-von-Humboldt Foundation (Germany);
ICSC (Italy); 
Severo Ochoa and Mar\'ia de Maeztu Units of Excellence, GVA, XuntaGal, GENCAT, InTalent-Inditex and Prog.~Atracci\'on Talento CM (Spain);
SRC (Sweden);
the Leverhulme Trust, the Royal Society and UKRI (United Kingdom).

\addcontentsline{toc}{section}{References}
\bibliographystyle{LHCb}
\bibliography{main,standard,smog,LHCb-PAPER,LHCb-CONF,LHCb-DP,LHCb-TDR}

\ifx\mcitethebibliography\mciteundefinedmacro
\PackageError{LHCb.bst}{mciteplus.sty has not been loaded}
{This bibstyle requires the use of the mciteplus package.}\fi
\providecommand{\href}[2]{#2}
\begin{mcitethebibliography}{10}
\mciteSetBstSublistMode{n}
\mciteSetBstMaxWidthForm{subitem}{\alph{mcitesubitemcount})}
\mciteSetBstSublistLabelBeginEnd{\mcitemaxwidthsubitemform\space}
{\relax}{\relax}

\bibitem{Salati:2010rc}
P.~Salati, F.~Donato, and N.~Fornengo,
  \ifthenelse{\boolean{articletitles}}{\emph{{Indirect Dark Matter Detection
  with Cosmic Antimatter}},
  }{}\href{http://arxiv.org/abs/1003.4124}{{\normalfont\ttfamily
  arXiv:1003.4124}}\relax
\mciteBstWouldAddEndPuncttrue
\mciteSetBstMidEndSepPunct{\mcitedefaultmidpunct}
{\mcitedefaultendpunct}{\mcitedefaultseppunct}\relax
\EndOfBibitem
\bibitem{AMS:2021nhj}
AMS collaboration, M.~Aguilar {\em et~al.},
  \ifthenelse{\boolean{articletitles}}{\emph{{The Alpha Magnetic Spectrometer
  (AMS) on the international space station: Part II \textemdash{} Results from
  the first seven years}},
  }{}\href{https://doi.org/10.1016/j.physrep.2020.09.003}{Phys.\ Rept.\
  \textbf{894} (2021) 1}\relax
\mciteBstWouldAddEndPuncttrue
\mciteSetBstMidEndSepPunct{\mcitedefaultmidpunct}
{\mcitedefaultendpunct}{\mcitedefaultseppunct}\relax
\EndOfBibitem
\bibitem{STingCERNseminar}
S.~Ting, \ifthenelse{\boolean{articletitles}}{\emph{{Latest results from AMS on
  the International Space Station}}, }{} 8 June 2023.
\newblock
  \href{https://indico.cern.ch/event/1275785/}{https://indico.cern.ch/event/1275785/}\relax
\mciteBstWouldAddEndPuncttrue
\mciteSetBstMidEndSepPunct{\mcitedefaultmidpunct}
{\mcitedefaultendpunct}{\mcitedefaultseppunct}\relax
\EndOfBibitem
\bibitem{LHCb-DP-2008-001}
LHCb collaboration, A.~A. Alves~Jr.\ {\em et~al.},
  \ifthenelse{\boolean{articletitles}}{\emph{{The \lhcb detector at the LHC}},
  }{}\href{https://doi.org/10.1088/1748-0221/3/08/S08005}{JINST \textbf{3}
  (2008) S08005}\relax
\mciteBstWouldAddEndPuncttrue
\mciteSetBstMidEndSepPunct{\mcitedefaultmidpunct}
{\mcitedefaultendpunct}{\mcitedefaultseppunct}\relax
\EndOfBibitem
\bibitem{LHCb-DP-2014-002}
LHCb collaboration, R.~Aaij {\em et~al.},
  \ifthenelse{\boolean{articletitles}}{\emph{{LHCb detector performance}},
  }{}\href{https://doi.org/10.1142/S0217751X15300227}{Int.\ J.\ Mod.\ Phys.\
  \textbf{A30} (2015) 1530022},
  \href{http://arxiv.org/abs/1412.6352}{{\normalfont\ttfamily
  arXiv:1412.6352}}\relax
\mciteBstWouldAddEndPuncttrue
\mciteSetBstMidEndSepPunct{\mcitedefaultmidpunct}
{\mcitedefaultendpunct}{\mcitedefaultseppunct}\relax
\EndOfBibitem
\bibitem{smog}
C.~Barschel, {\em {Precision luminosity measurement at LHCb with beam-gas
  imaging}}, PhD thesis, RWTH Aachen U., 2014,
  \href{https://cds.cern.ch/record/1693671}{CERN-THESIS-2013-301}\relax
\mciteBstWouldAddEndPuncttrue
\mciteSetBstMidEndSepPunct{\mcitedefaultmidpunct}
{\mcitedefaultendpunct}{\mcitedefaultseppunct}\relax
\EndOfBibitem
\bibitem{Bursche:2649878}
A.~Bursche {\em et~al.}, \ifthenelse{\boolean{articletitles}}{\emph{{Physics
  opportunities with the fixed-target program of the LHCb experiment using an
  unpolarized gas target}}, }{},
  \href{https://cds.cern.ch/record/2649878}{CERN-LHCb-PUB-2018-015}\relax
\mciteBstWouldAddEndPuncttrue
\mciteSetBstMidEndSepPunct{\mcitedefaultmidpunct}
{\mcitedefaultendpunct}{\mcitedefaultseppunct}\relax
\EndOfBibitem
\bibitem{LHCb-DP-2023-002}
LHCb collaboration, R.~Aaij {\em et~al.},
  \ifthenelse{\boolean{articletitles}}{\emph{{Helium identification with
  LHCb}}, }{}\href{https://doi.org/10.1088/1748-0221/19/02/P02010}{JINST
  \textbf{19} (2024) P02010},
  \href{http://arxiv.org/abs/2310.05864}{{\normalfont\ttfamily
  arXiv:2310.05864}}\relax
\mciteBstWouldAddEndPuncttrue
\mciteSetBstMidEndSepPunct{\mcitedefaultmidpunct}
{\mcitedefaultendpunct}{\mcitedefaultseppunct}\relax
\EndOfBibitem
\bibitem{Baker:2019fjx}
S.~K. Baker, {\em {Measurement of deuterons at LHCb}}, PhD thesis, Imperial
  Coll., London, 2019,
  doi:~\href{https://doi.org/10.25560/97672}{10.25560/97672}\relax
\mciteBstWouldAddEndPuncttrue
\mciteSetBstMidEndSepPunct{\mcitedefaultmidpunct}
{\mcitedefaultendpunct}{\mcitedefaultseppunct}\relax
\EndOfBibitem
\bibitem{LHCb-TDR-006}
LHCb collaboration, \ifthenelse{\boolean{articletitles}}{\emph{{LHCb outer
  tracker: Technical Design Report}}, }{}
  \href{http://cdsweb.cern.ch/search?p=CERN-LHCC-2001-024&f=reportnumber&action_search=Search&c=LHCb}
  {CERN-LHCC-2001-024}, 2001\relax
\mciteBstWouldAddEndPuncttrue
\mciteSetBstMidEndSepPunct{\mcitedefaultmidpunct}
{\mcitedefaultendpunct}{\mcitedefaultseppunct}\relax
\EndOfBibitem
\bibitem{LHCb-DP-2017-001}
P.~d'Argent {\em et~al.}, \ifthenelse{\boolean{articletitles}}{\emph{{Improved
  performance of the LHCb Outer Tracker in LHC Run 2}},
  }{}\href{https://doi.org/10.1088/1748-0221/12/11/P11016}{JINST \textbf{12}
  (2017) P11016}, \href{http://arxiv.org/abs/1708.00819}{{\normalfont\ttfamily
  arXiv:1708.00819}}\relax
\mciteBstWouldAddEndPuncttrue
\mciteSetBstMidEndSepPunct{\mcitedefaultmidpunct}
{\mcitedefaultendpunct}{\mcitedefaultseppunct}\relax
\EndOfBibitem
\bibitem{LHCb-DP-2012-002}
A.~A. Alves~Jr.\ {\em et~al.},
  \ifthenelse{\boolean{articletitles}}{\emph{{Performance of the LHCb muon
  system}}, }{}\href{https://doi.org/10.1088/1748-0221/8/02/P02022}{JINST
  \textbf{8} (2013) P02022},
  \href{http://arxiv.org/abs/1211.1346}{{\normalfont\ttfamily
  arXiv:1211.1346}}\relax
\mciteBstWouldAddEndPuncttrue
\mciteSetBstMidEndSepPunct{\mcitedefaultmidpunct}
{\mcitedefaultendpunct}{\mcitedefaultseppunct}\relax
\EndOfBibitem
\bibitem{DeCianThesis}
M.~De~Cian, {\em {Track reconstruction efficiency and analysis of $B^0 \to
  K^{*0} \mu^+ \mu^-$ at the \lhcb experiment}}, PhD thesis, Zurich University,
  2013, \url{https://cds.cern.ch/record/1605179}\relax
\mciteBstWouldAddEndPuncttrue
\mciteSetBstMidEndSepPunct{\mcitedefaultmidpunct}
{\mcitedefaultendpunct}{\mcitedefaultseppunct}\relax
\EndOfBibitem
\bibitem{PatForward}
O.~Callot and S.~Hansmann-Menzemer,
  \ifthenelse{\boolean{articletitles}}{\emph{{The forward tracking: Algorithm
  and performance studies}}}{},
  \href{https://cds.cern.ch/record/1033584}{{LHCb-2007-015,
  CERN-LHCb-2007-015}}, CERN, Geneva, 2007\relax
\mciteBstWouldAddEndPuncttrue
\mciteSetBstMidEndSepPunct{\mcitedefaultmidpunct}
{\mcitedefaultendpunct}{\mcitedefaultseppunct}\relax
\EndOfBibitem
\bibitem{Ostapchenko:2010vb}
S.~Ostapchenko, \ifthenelse{\boolean{articletitles}}{\emph{{Monte Carlo
  treatment of hadronic interactions in enhanced Pomeron scheme: QGSJET-II
  model}}, }{}\href{https://doi.org/10.1103/PhysRevD.83.014018}{Phys.\ Rev.\
  \textbf{D83} (2011) 014018},
  \href{http://arxiv.org/abs/1010.1869}{{\normalfont\ttfamily
  arXiv:1010.1869}}\relax
\mciteBstWouldAddEndPuncttrue
\mciteSetBstMidEndSepPunct{\mcitedefaultmidpunct}
{\mcitedefaultendpunct}{\mcitedefaultseppunct}\relax
\EndOfBibitem
\bibitem{Sjostrand:2006za}
T.~Sj\"{o}strand, S.~Mrenna, and P.~Skands,
  \ifthenelse{\boolean{articletitles}}{\emph{{PYTHIA 6.4 physics and manual}},
  }{}\href{https://doi.org/10.1088/1126-6708/2006/05/026}{JHEP \textbf{05}
  (2006) 026}, \href{http://arxiv.org/abs/hep-ph/0603175}{{\normalfont\ttfamily
  arXiv:hep-ph/0603175}}\relax
\mciteBstWouldAddEndPuncttrue
\mciteSetBstMidEndSepPunct{\mcitedefaultmidpunct}
{\mcitedefaultendpunct}{\mcitedefaultseppunct}\relax
\EndOfBibitem
\bibitem{ALICE-PUBLIC-2017-010}
ALICE collaboration, \ifthenelse{\boolean{articletitles}}{\emph{{Supplemental
  material: afterburner for generating light \mbox{(anti-)nuclei} with
  QCD-inspired event generators in pp collisions}}, }{},
  \href{https://cds.cern.ch/record/2285500}{ALICE-PUBLIC-2017-010}\relax
\mciteBstWouldAddEndPuncttrue
\mciteSetBstMidEndSepPunct{\mcitedefaultmidpunct}
{\mcitedefaultendpunct}{\mcitedefaultseppunct}\relax
\EndOfBibitem
\bibitem{Allison:2006ve}
Geant4 collaboration, J.~Allison {\em et~al.},
  \ifthenelse{\boolean{articletitles}}{\emph{{Geant4 developments and
  applications}}, }{}\href{https://doi.org/10.1109/TNS.2006.869826}{IEEE
  Trans.\ Nucl.\ Sci.\  \textbf{53} (2006) 270}\relax
\mciteBstWouldAddEndPuncttrue
\mciteSetBstMidEndSepPunct{\mcitedefaultmidpunct}
{\mcitedefaultendpunct}{\mcitedefaultseppunct}\relax
\EndOfBibitem
\bibitem{Agostinelli:2002hh}
Geant4 collaboration, S.~Agostinelli {\em et~al.},
  \ifthenelse{\boolean{articletitles}}{\emph{{Geant4: A simulation toolkit}},
  }{}\href{https://doi.org/10.1016/S0168-9002(03)01368-8}{Nucl.\ Instrum.\
  Meth.\  \textbf{A506} (2003) 250}\relax
\mciteBstWouldAddEndPuncttrue
\mciteSetBstMidEndSepPunct{\mcitedefaultmidpunct}
{\mcitedefaultendpunct}{\mcitedefaultseppunct}\relax
\EndOfBibitem
\bibitem{LHCb-PROC-2011-006}
M.~Clemencic {\em et~al.}, \ifthenelse{\boolean{articletitles}}{\emph{{The
  \lhcb simulation application, Gauss: Design, evolution and experience}},
  }{}\href{https://doi.org/10.1088/1742-6596/331/3/032023}{J.\ Phys.\ Conf.\
  Ser.\  \textbf{331} (2011) 032023}\relax
\mciteBstWouldAddEndPuncttrue
\mciteSetBstMidEndSepPunct{\mcitedefaultmidpunct}
{\mcitedefaultendpunct}{\mcitedefaultseppunct}\relax
\EndOfBibitem
\bibitem{Stahl:2017nlr}
LHCb collaboration, M.~Stahl,
  \ifthenelse{\boolean{articletitles}}{\emph{{Machine learning and parallelism
  in the reconstruction of LHCb and its upgrade}},
  }{}\href{https://doi.org/10.1088/1742-6596/898/4/042042}{J.\ Phys.\ Conf.\
  Ser.\  \textbf{898} (2017) 042042},
  \href{http://arxiv.org/abs/1710.08947}{{\normalfont\ttfamily
  arXiv:1710.08947}}\relax
\mciteBstWouldAddEndPuncttrue
\mciteSetBstMidEndSepPunct{\mcitedefaultmidpunct}
{\mcitedefaultendpunct}{\mcitedefaultseppunct}\relax
\EndOfBibitem
\bibitem{LHCb-DP-2019-001}
R.~Aaij {\em et~al.}, \ifthenelse{\boolean{articletitles}}{\emph{{Design and
  performance of the LHCb trigger and full real-time reconstruction in Run 2 of
  the LHC}}, }{}\href{https://doi.org/10.1088/1748-0221/14/04/P04013}{JINST
  \textbf{14} (2019) P04013},
  \href{http://arxiv.org/abs/1812.10790}{{\normalfont\ttfamily
  arXiv:1812.10790}}\relax
\mciteBstWouldAddEndPuncttrue
\mciteSetBstMidEndSepPunct{\mcitedefaultmidpunct}
{\mcitedefaultendpunct}{\mcitedefaultseppunct}\relax
\EndOfBibitem
\bibitem{PhDLucarelli}
C.~Lucarelli, {\em {Extending the physics reach of the fixed-target programme
  at the LHCb experiment}}, PhD thesis, Florence University, 2024,
  \url{https://cds.cern.ch/record/2889345}\relax
\mciteBstWouldAddEndPuncttrue
\mciteSetBstMidEndSepPunct{\mcitedefaultmidpunct}
{\mcitedefaultendpunct}{\mcitedefaultseppunct}\relax
\EndOfBibitem
\bibitem{Smirnov}
N.~Smirnov, \ifthenelse{\boolean{articletitles}}{\emph{{Table for estimating
  the goodness of fit of empirical distributions}},
  }{}\href{https://doi.org/10.1214/aoms/1177730256}{Ann.\ Math.\ Stat.\
  \textbf{19} (1948) 279}\relax
\mciteBstWouldAddEndPuncttrue
\mciteSetBstMidEndSepPunct{\mcitedefaultmidpunct}
{\mcitedefaultendpunct}{\mcitedefaultseppunct}\relax
\EndOfBibitem
\bibitem{Keras}
F.~Chollet {\em et~al.}, \ifthenelse{\boolean{articletitles}}{\emph{Keras}, }{}
  \url{https://keras.io}, 2015\relax
\mciteBstWouldAddEndPuncttrue
\mciteSetBstMidEndSepPunct{\mcitedefaultmidpunct}
{\mcitedefaultendpunct}{\mcitedefaultseppunct}\relax
\EndOfBibitem
\bibitem{TF}
M.~Abadi {\em et~al.}, \ifthenelse{\boolean{articletitles}}{\emph{{TensorFlow}:
  Large-scale machine learning on heterogeneous systems}, }{}
  \url{https://www.tensorflow.org/}, 2015\relax
\mciteBstWouldAddEndPuncttrue
\mciteSetBstMidEndSepPunct{\mcitedefaultmidpunct}
{\mcitedefaultendpunct}{\mcitedefaultseppunct}\relax
\EndOfBibitem
\bibitem{LHCb-PAPER-2018-031}
LHCb collaboration, R.~Aaij {\em et~al.},
  \ifthenelse{\boolean{articletitles}}{\emph{{Measurement of antiproton
  production in \proton{}He collisions at $\sqsnn=110$ GeV}},
  }{}\href{https://doi.org/10.1103/PhysRevLett.121.222001}{Phys.\ Rev.\ Lett.\
  \textbf{121} (2018) 222001},
  \href{http://arxiv.org/abs/1808.06127}{{\normalfont\ttfamily
  arXiv:1808.06127}}\relax
\mciteBstWouldAddEndPuncttrue
\mciteSetBstMidEndSepPunct{\mcitedefaultmidpunct}
{\mcitedefaultendpunct}{\mcitedefaultseppunct}\relax
\EndOfBibitem
\bibitem{Podolanski01011954}
J.~Podolanski and R.~Armenteros,
  \ifthenelse{\boolean{articletitles}}{\emph{{Analysis of V-events}},
  }{}\href{https://doi.org/10.1080/14786440108520416}{Philos.\ Mag.\
  \textbf{45} (1954) 13}\relax
\mciteBstWouldAddEndPuncttrue
\mciteSetBstMidEndSepPunct{\mcitedefaultmidpunct}
{\mcitedefaultendpunct}{\mcitedefaultseppunct}\relax
\EndOfBibitem
\end{mcitethebibliography}

\newpage
\centerline
{\large\bf LHCb collaboration}
\begin
{flushleft}
\small
R.~Aaij$^{38}$\lhcborcid{0000-0003-0533-1952},
A.S.W.~Abdelmotteleb$^{57}$\lhcborcid{0000-0001-7905-0542},
C.~Abellan~Beteta$^{51}$\lhcborcid{0009-0009-0869-6798},
F.~Abudin{\'e}n$^{57}$\lhcborcid{0000-0002-6737-3528},
T.~Ackernley$^{61}$\lhcborcid{0000-0002-5951-3498},
A. A. ~Adefisoye$^{69}$\lhcborcid{0000-0003-2448-1550},
B.~Adeva$^{47}$\lhcborcid{0000-0001-9756-3712},
M.~Adinolfi$^{55}$\lhcborcid{0000-0002-1326-1264},
P.~Adlarson$^{85}$\lhcborcid{0000-0001-6280-3851},
C.~Agapopoulou$^{14}$\lhcborcid{0000-0002-2368-0147}
% ...

{\footnotesize \it

% ...
$^{14}$Universit{\'e} Paris-Saclay, CNRS/IN2P3, IJCLab, Orsay, France\\
% ...
$^{38}$Nikhef National Institute for Subatomic Physics, Amsterdam, Netherlands\\
% ...
$^{47}$Instituto Galego de F{\'\i}sica de Altas Enerx{\'\i}as (IGFAE), Universidade de Santiago de Compostela, Santiago de Compostela, Spain\\
% ...
$^{51}$Physik-Institut, Universit{\"a}t Z{\"u}rich, Z{\"u}rich, Switzerland\\
% ...
$^{55}$H.H. Wills Physics Laboratory, University of Bristol, Bristol, United Kingdom\\
% ...
$^{57}$Department of Physics, University of Warwick, Coventry, United Kingdom\\
% ...
$^{60}$School of Physics and Astronomy, University of Glasgow, Glasgow, United Kingdom\\
$^{61}$Oliver Lodge Laboratory, University of Liverpool, Liverpool, United Kingdom\\
% ...
$^{69}$Syracuse University, Syracuse, NY, United States\\
% ...
$^{85}$Department of Physics and Astronomy, Uppsala University, Uppsala, Sweden, associated to $^{60}$\\
% ...
}
\end{flushleft}

\end{document}